\documentclass{WileyChemistry-template}

  \definecolor{pink}{HTML}{A03472}
  \definecolor{dlr_blue}{HTML}{00658b}
  \definecolor{dlr_darkblue}{HTML}{003245}
  \definecolor{dlr_green}{HTML}{94b333}
  \definecolor{dlr_yellow}{HTML}{efca0a}
  \definecolor{frenchblue}{HTML}{318ce7}
  \definecolor{fauxfern}{HTML}{4f7949}
  \definecolor{fbpink}{HTML}{E7318C}
  \definecolor{fborange}{HTML}{E78C31}
  \definecolor{mustard}{HTML}{FFEE6B}

\DeclareSIUnit\amperehour{Ah}
\usepackage[minimal, modules=reactions]{chemmacros}
    \chemsetup{
      greek=upgreek,
    }
\usepackage{arydshln}
\usepackage{nicefrac}
\usepackage{microtype}
\usepackage[textsize=scriptsize]{todonotes}
\newcommand{\secref}[2]{\hyperref[#1]{#2}}

\title{\raggedright Modelling and Simulation of an Alkaline Ni/Zn Cell$^\dag$}

\author{
\begin{minipage}{\textwidth}
	Felix K. Schwab,*\textsuperscript{\texttt{+},[a,b]}
	Britta Doppl,\textsuperscript{\texttt{+},[a,b]}
	Niklas J. Herrmann,\textsuperscript{[a,b]}
	Alice Boudet,\textsuperscript{[d]}
	Shadi Mirhashemi,\textsuperscript{[d]}
	Sylvain Brimaud,\textsuperscript{[e]}
	Birger Horstmann*\textsuperscript{,[a,b,c]}
\end{minipage}
}

\newcommand{\affiliation}{
\begin{itemize}

\item[{[a]}] Dr. F. K. Schwab*\textsuperscript{\texttt{+}}, B. Doppl\textsuperscript{\texttt{+}}, N. J. Herrmann, Prof. Dr. B. Horstmann*\\
Institute of Engineering Thermodynamics - Computational Electrochemistry, German Aerospace Center (DLR), Wilhelm-Runge-Str. 10, 89081 Ulm, Germany \\
E-mail: felix.schwab@dlr.de, birger.horstmann@dlr.de

\item[{[b]}] Dr. F. K. Schwab*\textsuperscript{\texttt{+}}, B. Doppl\textsuperscript{\texttt{+}}, N. J. Herrmann, Prof. Dr. B. Horstmann*\\
Helmholtz Institute Ulm (HIU), Helmholtzstr. 11, 89081 Ulm, Germany

\item[{[c]}] Prof. Dr. B. Horstmann*\\
Faculty of Natural Science, Ulm University, Albert-Einstein-Allee 11, 89069 Ulm, Germany

\item[{[d]}] Dr. Alice Boudet, Dr. Shadi Mirhashemi\\
SunErgy, 85-93 Boulevard Alsace-Lorraine, 93115 Rosny-sous-Bois, France

\item[{[e]}] Dr. Sylvain Brimaud\\
Zentrum für Sonnenenergie- und Wasserstoff-Forschung Baden-Württemberg (ZSW), Helmholtzstr. 8, 89081 Ulm, Germany

\item[{[\texttt{*}]}] These are corresponding authors.
\item[{[\texttt{+}]}] These authors contributed equally.
\item[{[\texttt{\dag}]}] Electronic Supplementary Information (ESI) available.

\end{itemize}
}

\renewcommand{\dedication}{
	\begin{minipage}{\textwidth}
	\end{minipage}
}

\renewcommand{\abstract}{%
Nickel/zinc (\ce{Ni}/\ce{Zn}) technology is a promising post-lithium battery type for stationary applications with respect to aspects such as safety, environmental compatibility and resource availability.
Although this battery type has been known for a long time,
the theoretical knowledge about the processes taking place in the battery is limited.
In order to gain a deeper understanding of the general cycling behaviour and the underlying processes,
but also specific phenomena intrinsic to zinc-based cells such as zinc shape change,
we carry out simulations based on a thermodynamically consistent and volume-averaged continuum model.
We use a \ce{Ni}/\ce{Zn} prototype cell as a reference framework to provide a basis for modelling, parameter estimation and systematic comparison between simulated and experimental cell behaviour to improve cyclability and performance.
}

\begin{document}

\twocolumn[\vspace{-1.5cm}\maketitle\vspace{-1cm}
	\textit{\dedication}\vspace{0.4cm}]
\small{\begin{shaded}
		\noindent\abstract
	\end{shaded}
}

\begin{figure} [!b]
\begin{minipage}[t]{\columnwidth}{\rule{\columnwidth}{1pt}\footnotesize{\textsf{\affiliation}}}\end{minipage}
\end{figure}


%
%
%
\section*{Introduction}
\label{sec:introduction}
  %
  Zinc metal batteries come with a beneficial set of properties,
  which - in principle - enables them to compete with the nowadays wide-spread lithium-ion batteries,
  especially in stationary applications.
  There,
  when weight is not the only criteria for a successful utilisation,
  zinc electrodes convince with a rather high theoretical capacity (~\SI{819.7}{\milli\amperehour\per\gram}~)
  and relative stability against corrosion in aqueous electrolytes.\cite{Clark2020}
  The usage of such electrolytes significantly lowers their environmental harm, toxicity and flammability,
  which predestines them to areas where high safety regularities apply,
  while their conductivity is high.\cite{Borchers2021,Liang2023}
  Both, zinc and aqueous electrolytes, are typically low-cost.
  \par
  %
  These appealing characteristics reflect in a long-standing and ongoing interest in this technology,
  which has resulted in a whole family of zinc-metal batteries,\cite{Zhang2009,Reddy2010,Clark2020,Borchers2021}
  e.g. \ce{Zn}/\ce{Ni}, \ce{Zn}/Air, \ce{Zn}/\ce{MnO2} or \ce{Zn}/\ce{Ag}, amongst others.
  But despite this continuous research effort,\cite{Herrmann2024,Herrmann2024b}
  certain processes such as the redistribution of zinc -- the so-called shape change,
  densification and passivation,
  dendrite formation or hydrogen formation persist,\cite{McLarnon1991,Reddy2010,Mainar2016,Lu2021,Naveed2022}
  regardless of a better understanding and progress with theses topics.
  %
  %
  \par
  %
  A promising electrode pair is \ce{Zn}/\ce{Ni} (in the following the more usual term \ce{Ni}/\ce{Zn} is used),
  which makes use of the well-established $\beta$-\ce{NiO(OH)}/$\beta$-\ce{Ni(OH)2} insertion material as the positive electrode that is stable in alkaline media.\cite{Shukla2001}
  When this element combination is constructed as a vented battery,
  high energy and power densities\cite{Shukla2001,Huot2009,Bogomolov2024} and cycle life at good depths of discharge (DoD)\cite{Reisner1989,Turney2017} may be achieved.
  The theoretical capacity of the positive electrode (~\SI{289.1}{\milli\amperehour\per\gram}~)
  is the limit for the cell,\cite{Shukla2001}
  which may be maximally utilised to up to around 90\% due to oxygen formation at the end of charge.\cite{Jindra1992,McBreen1994}
  %
  \par
  %
  Along with the history of zinc metal batteries,
  efforts to model them on a continuum scale date back a long time.
  Numerous modelling works can be found for the stand-alone zinc electrode, or in any combination
  (see Ref.\cite{Podlaha1994,Wen2022,Herrmann2024},
  amongst others).
  Important topics have been the correct description of the \ce{Zn}/\ce{ZnO} dissolution-precipitation reaction,\cite{Choi1976,Sunu1980,Chen1993}
  hydrogen formation,\cite{Deiss2002}
  the change in active surface area due to zinc conversion and hindered transport through the \ce{ZnO} porous layer.\cite{Mao1992,Chen1993,Stamm2017,Schmitt2019}
  Equally,
  the $\beta$-\ce{NiO(OH)}/$\beta$-\ce{Ni(OH)2} electrode has been extensively modelled,
  mostly in the context of \ce{Ni}/\ce{Cd} or metal-hydride batteries,
  see Ref.\cite{Fan1991,Paxton1997},
  amongst others.
  Here, the electrode's OCP curve and hysteresis,\cite{Gu1999,Pan2002,Albertus2008} the proton insertion modelled as solid diffusion\cite{Mao1994,DeVidts1995} and the oxygen formation at the end of charge\cite{Gu1999,Pan2002,Albertus2008} have been of importance.
  In the case of \ce{Ni}/\ce{Zn} continuum models only a few works exist.
  In an early one, Choi and Yao\cite{Choi1979} examined the cell type with a focus on the \ce{Ni} electrode,
  already including the convection of the electrolyte solution but neglecting the \ce{O2} formation.
  The influence of operation conditions,
  e.g. the rate of charge and species concentrations,
  on the charge acceptance of the electrode is described.
  Miller \textit{et~al.}\cite{Miller1988} analysed concentration levels in the whole extent of the cell.
  Isaacson \textit{et~al.}\cite{Isaacson1990} used a 2D continuum model to investigate concentration and current density distribution, the zinc shape change and the influence of an electrolyte reservoir.
  Again, \ce{O2} formation and other aspects of the \ce{Ni} electrode are not part of the model due to an emphasis on the \ce{Zn} electrode.
  Later, Arise \textit{et~al.}\cite{Arise2010,Arise2013} used a simplified model to study transient concentration profiles in the cell and to explain experimentally found morphology changes in the \ce{Zn} electrode,
  without e.g. accounting for the zinc dissolution-precipitation reactions.
  Huang \textit{et~al.}\cite{Huang2017} used a continuum model to assist the development of new nickel high power electrodes.
  The model includes \ce{O2} formation and solid diffusion proton insertion. 
  %
  %
  \par
  The aim of this study is to present a 3D model of a \ce{Ni}/\ce{Zn} cell,
  which is able to represent the experimental data of such a cell prototype.
  For this purpose,
  we combine the relevant aspects of the \ce{Ni}/\ce{Zn} technology in one continuum model,
  focusing on dissolution-precipitation at the \ce{Zn} electrode as well as proton insertion and \ce{O2} formation at the \ce{Ni} electrode.
  Concentrated solution theory for the electrolyte solution including convection completes the model.
  On this basis,
  we discuss, evaluate, and compare the predictions of the cell model to experimental results and images.
  The modelling is based on the work of Latz and Zausch,\cite{Latz2011,Latz2015}
  who introduced a thermodynamically consistent transport theory for lithium ion batteries.
  Aspects of the \ce{Zn} electrode are mainly based on works of
  Stamm \textit{et~al.} and Schmitt \textit{et~al.},\cite{Stamm2017,Schmitt2019}
  while the model for the \ce{Ni} electrode is largely inspired by preliminary work by
  Doppl\cite{Doppl2024}, which is based on
  Paxton \textit{et~al.} and Albertus \textit{et~al.}\cite{Paxton1997,Albertus2008}
  \par
  Before stating our computational and experimental methods in \secref{sec:model}{the upcoming section},
  we formulate the reactions which drive the alkaline \ce{Ni}/\ce{Zn} cell 
  and describe the battery's working principle.
  %
  %
  \paragraph{Main working principle and reactions}
  \label{para:theory_reactions}
    \begin{figure}[tbp]
      \centering
        \includegraphics[width=8.4cm]{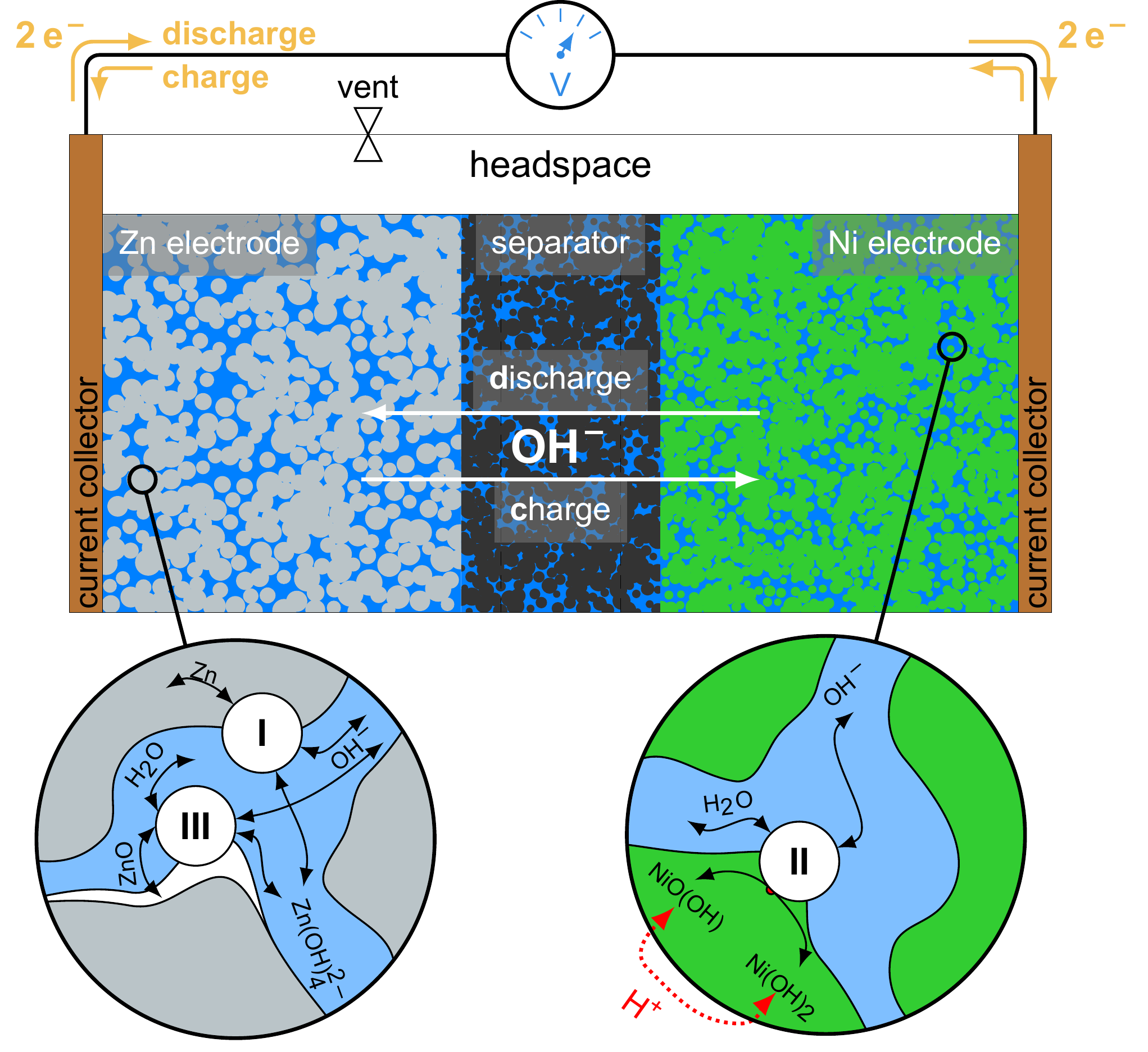}
        \caption{Schematic representation of a vented Ni/Zn cell together with the most important reactions.}
        \label{fig:nizn_cell}
    \end{figure}
    When discharging the \ce{Ni}/\ce{Zn} cell,
    the following happens:
    Firstly,
    at the negative \ce{Zn}/\ce{ZnO} electrode,
    zinc is dissolved and released into the strongly alkaline \ce{ZnO}-saturated \ce{KOH} electrolyte solution as \ce{Zn(OH)4^2-} (zincate, \Cref{fig:nizn_cell}\textcircled{\tiny{I}}) while providing two electrons,
    \begin{reaction}\label{rxn:elchem_reactions_I}
      Zn + 4 OH- <=> Zn(OH)4^{2-} + 2 e- \text{,}
    \end{reaction}
    with $\Delta\phi^{\circ} = -\SI{1.20}{\volt}$/SHE.
    Continued dissolution and re-deposition of \ce{Zn} changes the shape of the electrode.
    As seen in \Cref{rxn:elchem_reactions_I},
    this process requires \ce{OH-} (hydroxide) ions,
    which are transported through the separator coming from the positive $\beta$-\ce{NiO(OH)}/$\beta$-\ce{Ni(OH)2} electrode.
    Secondly,
    the electrons stemming from the zinc dissolution arrive through the electric circuit and split the water in the electrolyte solution (\Cref{fig:nizn_cell}\textcircled{\tiny{II}}).
    The emerging protons are inserted into the \ce{NiO(OH)} structure,
    \begin{reaction}\label{rxn:elchem_reactions_II}
      \beta-NiO(OH) + H2O + e- <=> \beta-Ni(OH)2 + OH- \text{,}
    \end{reaction}
    with $\Delta\phi^{\circ} = +\SI{0.49}{\volt}$/SHE,
    and the \ce{OH-} ions transported to the negative electrode close the circuit.
    Lastly,
    when the zincate concentration in the electrolyte solution rises above its saturation limit,
    a precipitation reaction takes place forming a \ce{ZnO} layer on top of the \ce{Zn} in the negative electrode (\Cref{fig:nizn_cell}\textcircled{\tiny{III}}),
    \begin{reaction}
      \label{rxn:chem_reaction}
      Zn(OH)4^{2-} <=> ZnO + H2O + 2 OH- \text{.}
    \end{reaction}
    This is an important part of the battery concept,
    as the precipitation allows \Cref{rxn:elchem_reactions_I} to continue efficiently through \ce{Zn(OH)4^{2-}} being stored compactly as \ce{ZnO}.
    However,
    this precipitation often forms a passivating layer on \ce{Zn} surfaces.
    When charging,
    all these processes are reversed.
    Overall,
    the cell reaction is
    \begin{reaction*}
      \label{rxn:total_reaction}
      Zn + 2 \beta-NiO(OH) + H2O <=> ZnO + 2 \beta-Ni(OH)2 \text{,}
    \end{reaction*}
    with $\Delta\phi^{\circ} = +\SI{1.69}{\volt}$/SHE.
    \par
    Due to thecell voltage,
    which lies outside the electrochemical stability window of water,
    $\Delta E = \SI{1.23}{\volt}$,
    two side reactions,
    hydrogen and oxygen evolution reactions (HER and OER),
    play a major role:
    \begin{reaction}
      \label{rxn:side_reactions_I}
      2H2O + 2e- <=> H2_{(aq)} + 2OH- \text{,}
    \end{reaction}
    \begin{reaction}
      \label{rxn:side_reactions_II}
      4OH- <=> 2H2O + O2_{(aq)} + 4e- \text{,}
    \end{reaction}
    with $\Delta\phi^{\circ} = +\SI{0.83}{\volt}$/SHE and $\Delta\phi^{\circ} = -\SI{0.40}{\volt}$/SHE,
    respectively.
    The main consequence of the OER is a reduced Coulombic efficiency,
    which in turn leads to more \ce{ZnO} being dissolved during charging than is precipitated during discharging.
    However,
    this \ce{ZnO} imbalance is counteracted by a portion of the \ce{O2} produced being transported from the \ce{Ni} electrode through the separator to the \ce{Zn} electrode,
    where it reacts to form \ce{ZnO},\cite{Jindra1992,Reddy2010} 
    \begin{reaction}
      \label{rxn:chem_reaction_II}
      2Zn + O2_{(aq)} <=> 2ZnO \text{.}
    \end{reaction}
    Excess \ce{O2} and \ce{H2} is released from the electrolyte through Reactions $\{7\}$ and $\{8\}$, see Section 9 of the ESI\dag.
    In such cells,
    but disregarded in this work,
    these gases are collected in a headspace (see \Cref{fig:nizn_cell}),
    which serves additionally as an electrolyte reservoir.
    There,
    the gases recombine to water at a catalyst or vent through a valve when pressure becomes critical.\cite{McLarnon1991,Jindra1992,McBreen1994,Shukla2001,Reddy2010}
    This causes loss of electrolyte solution,
    which may result in a so-called dry-out\cite{Reddy2010} leaving parts of the electrodes non-wetted.
    Occasional electrolyte refilling may be required.
  \par
  
  %
  %
  \paragraph{Structure}
  \label{para:structure}
  In the following,
  we will give details on the \secref{sec:model}{computational and experimental methods} used,
  pointing out the important aspects of the transport and reaction modelling.
  Then,
  in the \secref{sec:results_and_discussion}{results and discussion section},
  we present findings from the simulation of single cycles and long term cycling.
  They are compared to experimental results, discussed and optimisation strategies are outlined.
  Lastly,
  we give a \secref{sec:conclusion}{summary and conclusion} of our work.
\section*{Computational and Experimental Methods}
\label{sec:model}
  We conduct experiments and computational studies on the basis of a vented \ce{Ni}/\ce{Zn} cell with a strongly alkaline \ce{KOH} electrolyte solution.
  It hence combines a conversion-type \ce{Zn} electrode,
  which dissolves during the discharge process,
  with an insertion-type \ce{Ni} electrode,
  which stores protons.
  Due to the use of an aqueous electrolyte solution and operation outside of the electrochemical stability window of water,
  gases are produced affecting battery functioning.
  \par
  Regarding the cell model,
  the most important assumptions and approaches are derived from the description in the introductory \secref{para:theory_reactions}{paragraph on working principles and reactions},
  which will only be discussed in essence here.
  For further details,
  we refer the interested reader directly to the ESI\dag.
  \subsection*{Modelling approach}
  \label{subsec:theory_modelling}
    The main functionality of a battery cell,
    the conversion of chemical to electric energy through electrochemical reactions,
    happens at the interfaces between the solid electrodes and an electrolyte solution.
    Furthermore,
    the transport of electrons through the former and ions through the latter has to be ensured to close the circuit.
    In order to represent this behaviour in a consistent physico-chemical model,
    a physics-based model approach is chosen that includes the transport process in the form of balance equations and reactions as (electro-)chemical kinetics.
    Based on similar and already existing models for lithium-based or zinc-based batteries,
    the \ce{Ni}/\ce{Zn} cell discussed here is modelled.\cite{Latz2011,Latz2015,Stamm2017,Schmitt2019}
    In the following,
    only the basic assumptions and foundation of our model will be introduced.
    We will not list and explain the numerous model equations and parameters in detail here,
    but refer the interested reader to the ESI\dag.
    \paragraph{Transport in electrolyte and electrodes}
    \label{para:transport}
    Two crucial assumptions form the basis of our transport description.
    Firstly,
    the strongly alkaline electrolyte solution requires the use of concentrated solution theory,
    which introduces transport parameters that are independent of each other.\cite{Stamm2017}
    For each species $k$ in the electrolyte solution we have a diffusion coefficient $D^k$,
    a transference number $t^k$,
    and for the solution in total an ionic conductivity $\kappa$.
    Secondly,
    we use a volume-averaged description to circumvent a detailed representation of the electrodes' and the separator's micro-structure,\cite{Schmitt2019}
    which helps to make a three-dimensional simulation of a whole battery cell feasible.
    Volume averaging condenses the micro-structure information of a phase $\alpha$,
    e.g. solid electrode phase or liquid electrolyte phase,
    into two phase-specific quantities,
    volume fraction $\varepsilon^\alpha$ and tortuosity $\tau^\alpha$.
    In each point,
    the volume fractions of all solids, liquids and gases have to follow the sum constraint $\varepsilon^{\text{s}} + \varepsilon^{\text{l}} + \varepsilon^{\text{g}} = 1$.
    On an abstract level,
    this leads then to effective parameters and quantities,
    e.g. effective diffusion coefficients $D^{k,\text{eff.}} = \nicefrac{\varepsilon}{\tau^2} D^k$ or effective concentrations $c^{k,\text{eff.}} = \varepsilon c^k$,
    which have the micro-structure information inscribed.
    \par
    Within this framework,
    a major part are the transport or balance laws in each phase.
    In the liquid electrolyte phase,
    these transport equations comprise fluxes due to diffusion,
    electromigration and volume-centred convection.
    While diffusion and electromigration are caused by gradients in concentrations and electric potential,
    respectively,
    convection in our model roots from changes in the solid and liquid volume fractions inducing a pressure in the liquid, $p^\text{l}$,
    causing a displacement of liquid.\cite{Schmitt2019}
    Here, the volume-centred description\cite{Schammer2021,Kilchert2023} is used instead of the mass-centred one to facilitate the evaluation of the incompressibility condition.
    These coupled fluxes then drive the transport of species in the electrolyte solution,
    which are represented by local concentrations $c^k$, $k = \{ \ce{OH^-}, \ce{Zn(OH)4^{2-}}, \ce{O2_{(aq)}}, \ce{H2_{(aq)}} \}$.
    Furthermore,
    we solve for the electrolyte potential $\phi^\text{e}$.
    In the solid phases of the electrodes,
    only electronic conduction plays a role,
    which is caused by the gradient of the solid phase (electrode) potential $\phi^\text{s}$.
    \par
\begin{figure}[tbp]
      \centering
        \includegraphics[width=8.4cm]{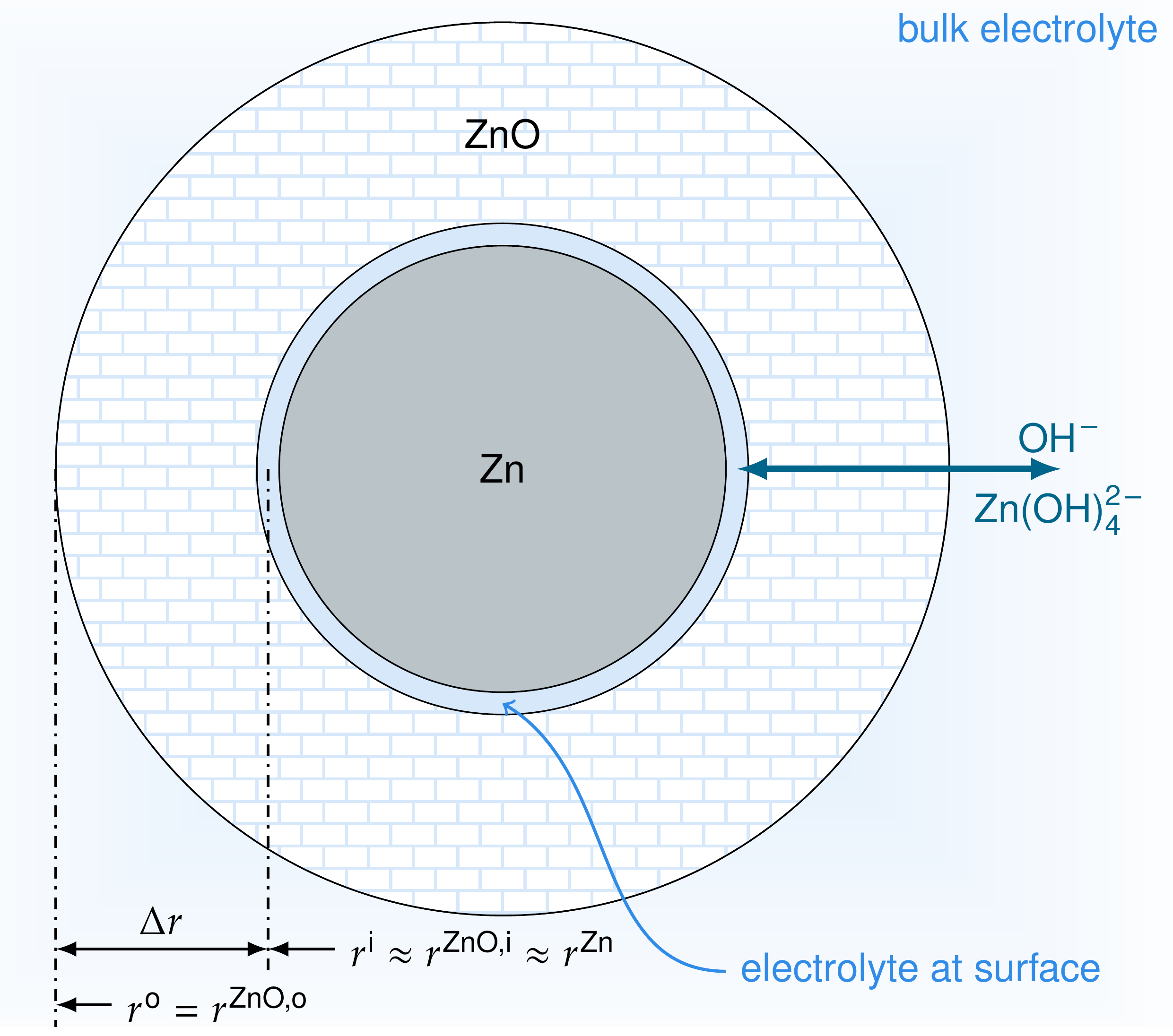}
        \caption{Illustration of a zinc particle with surrounding zinc oxide layer.
                 The layer slows down species transport to the zinc surface, influencing the respective reactions.
                 We assume that the inner radius of the \ce{ZnO} shell coincides with the outer radius of the \ce{Zn} particle.}
        \label{fig:surface_concentration}
    \end{figure}
\begin{figure}[htbp]
      \centering
        \includegraphics[width=5.73cm]{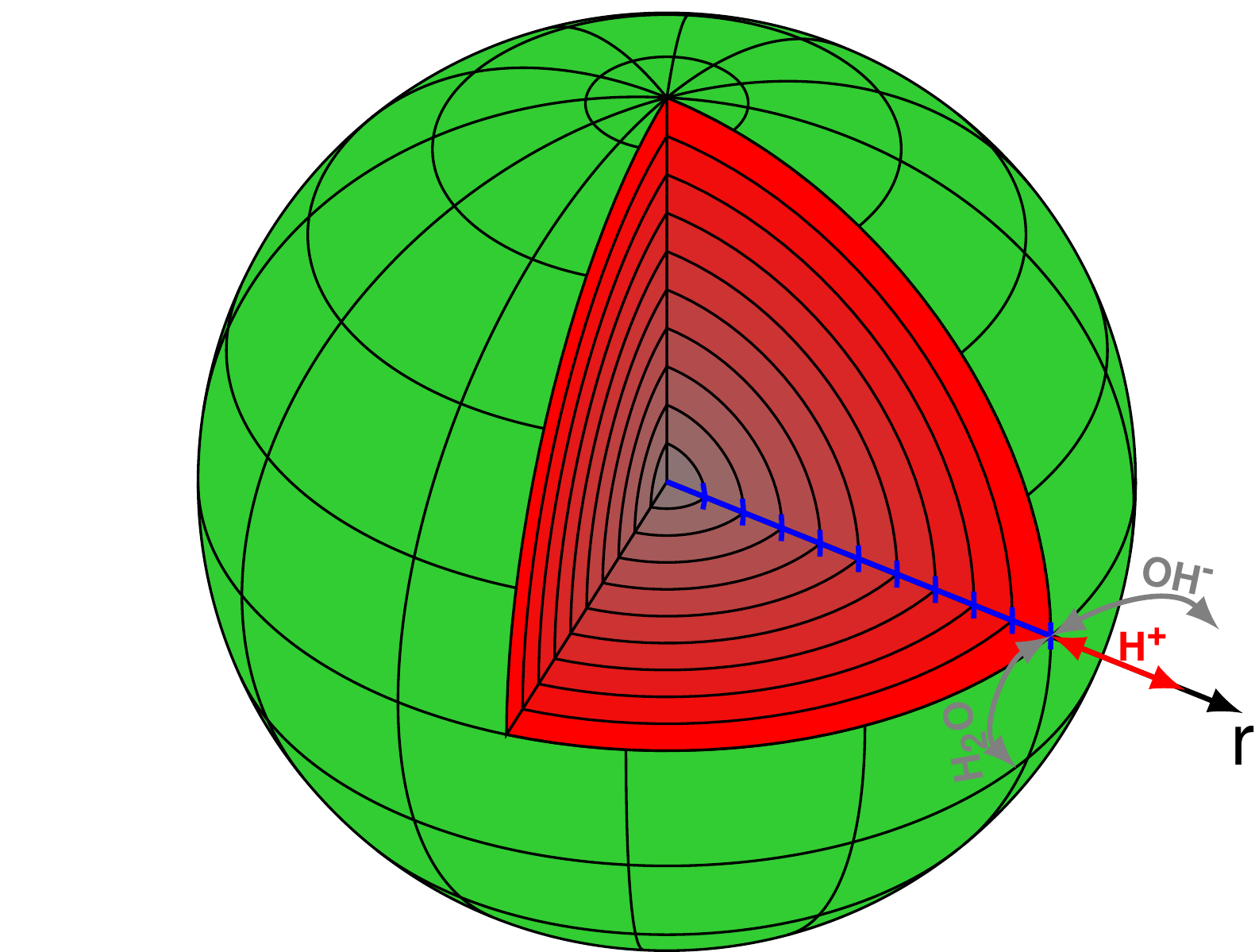}
        \caption{Depiction of a \ce{NiO(OH)}/\ce{Ni(OH)2} particle as a perfect sphere.
                 The protons entering the particle on the surface through \Cref{rxn:elchem_reactions_II} are inserted via a solid diffusion mechanism.}
        \label{fig:solid_diffusion}
    \end{figure}

    The difference between the latter and the former potential drives electrochemical reactions at the electrode-electrolyte interfaces,
    which are modelled as phenomenological Butler-Volmer equations.
    If \ce{O2_{(aq)}} and \ce{H2_{(aq)}} oversaturate in the electrolyte solution,
    they outgas and do not further influence the \ce{Ni}/\ce{Zn} cell.
    \paragraph{Zinc electrode}
    \label{para:zn_electrode}
    Both,
    the dissolution of the \ce{Zn} electrode due to \Cref{rxn:elchem_reactions_I}
    and the precipitation of \ce{ZnO} due to \Cref{rxn:chem_reaction} as soon as the electrolyte solution is saturated with zincate,
    result in a change of the respective volume fractions,
    $\varepsilon^{\ce{Zn}}$ and $\varepsilon^{\ce{ZnO}}$,
    modelled by balance equations.
    The latter reaction creates a \ce{ZnO} shell around the \ce{Zn} particles which acts as a passivation:\cite{Stamm2017}
    If present,
    the transport of \ce{OH^-} and \ce{Zn(OH)4^2-} to and from the zinc surface is hindered (see \Cref{fig:surface_concentration}).
    By calculating surface concentrations for these two species,
    $c^{k,\text{s}}$, $k = \{ \ce{OH^-}, \ce{Zn(OH)4^{2-}} \}$,
    this aspect of the zinc electrode enters and influences \Cref{rxn:elchem_reactions_I} and \Cref{rxn:side_reactions_I}.
    From a modelling point of view this is written as a balance between the hindered diffusion through the \ce{ZnO} shell and the reactions depending on species $k$ happening on the \ce{Zn} surface,
    \begin{equation}
      \left(1 - \varepsilon^{\ce{ZnO}\text{-L}} \right)^{\beta^{\ce{ZnO}\text{-L}}} D^{k} \frac{r^{\ce{ZnO,\text{o}}}}{r^{\ce{ZnO,\text{i}}}} \frac{\left( c^{k,\text{s}} - c^{k} \right)}{\Delta r} = \sum_r \nu^{kr} j^{r} \text{,}
      \label{eq:surface_concentration}
    \end{equation}
    where $\varepsilon^{\ce{ZnO}\text{-L}}$ and $\beta^{\ce{ZnO}\text{-L}}$ are the volume fraction and the Bruggeman coefficient\cite{Tjaden2016} of the \ce{ZnO} layer,
    converting the diffusion coefficient $D^{k}$ to an effective property.
    While the different radii correspond to those depicted in \Cref{fig:surface_concentration},
    the right-hand-side of the balance equation contains the sum over all reactions $r$ producing or consuming species $k$ via the reaction rate $j^{r}$ and the corresponding stoichiometric coefficient $\nu^{kr}$.
    Furthermore,
    oxygen recombination (\Cref{rxn:chem_reaction_II}) takes place at the \ce{Zn} electrode,
    which is modelled by a simple chemical reaction equation.
    \paragraph{Nickel electrode}
    \label{para:ni_electrode}
    In contrast,
    at the \ce{Ni} electrode,
    an insertion process takes place where protons enter the active material through the electrochemical \Cref{rxn:elchem_reactions_II}.
    From the surface they diffuse into the particle and are stored by forming \ce{Ni(OH)2},
    see \Cref{fig:solid_diffusion}.
    Assuming the \ce{NiO(OH)} particles are perfect spheres,
    we describe the storage process as a solid diffusion process of the proton concentration $c^{\ce{H^+}}$,
    \begin{equation}
      \frac{\partial c^{\ce{H^+}}}{\partial t} = \frac{1}{r^2} \frac{\partial}{\partial r} \left( D^{\ce{H^+}} r^2 \frac{\partial c^{\ce{H^+}}}{\partial r} \right) \text{,}
      \label{eq:radial_solid_diffusion}
    \end{equation}
    with $r$ being the radial coordinate in the particle and $D^{\ce{H^+}}$ being the solid diffusion coefficient of the protons (see Equation 7 of the ESI\dag).
    Since \ce{NiO(OH)} and \ce{Ni(OH)2} show a difference in mass density,
    the volume fraction of the active material in the \ce{Ni} electrode would change as in the \ce{Zn} electrode.
    As this difference is small,
    we neglect this effect here.
    %
    %
    %
    %
  %
  %
  %
  \subsection*{Simulation and setup}
  \label{subsec:setup}
\begin{figure}[htbp]
      \centering
        \includegraphics[width=8.4cm]{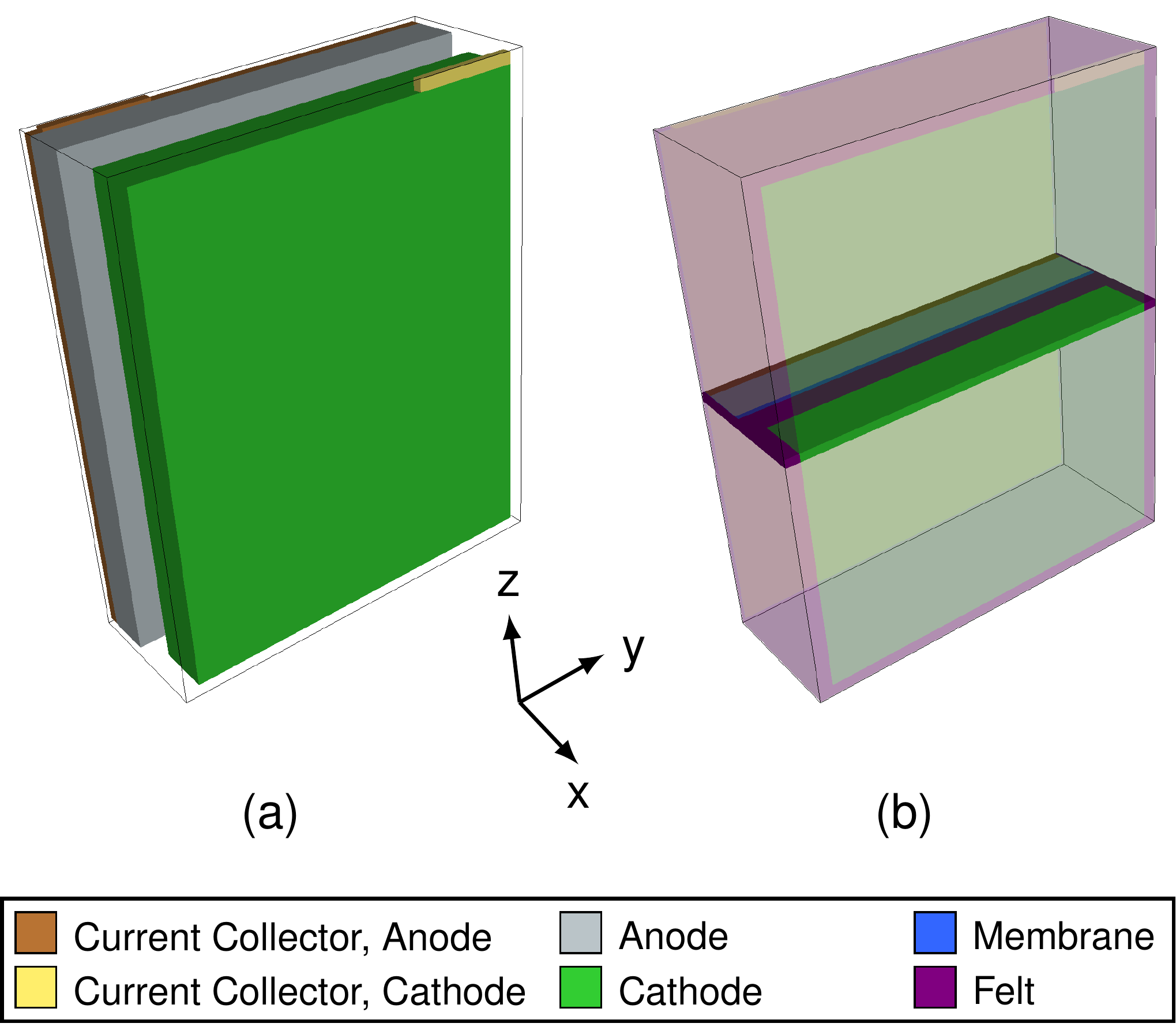}
        \caption{Setup of the simulation boxes.
                 (a) Detailed full cell setup, membrane and felt not shown.
                 (b) Thin, horizontal slice of the full cell serves as a reduced setup.
                 The x-axis has been scaled by a factor of 10 for a better depiction.}
        \label{fig:setup}
    \end{figure}

    The geometry of the simulation setup used in this work follows that of \textsc{SunErgy}'s prototype \ce{Ni}/\ce{Zn} cell as described in Section 2 of the ESI\dag.
    This gives a simulation box for the total cell as it is depicted in \Cref{fig:setup}a,
    which is used in a resolution of $13\times21\times23$ voxels for simulations where a three-dimensionally resolved structure is needed.
    For all other simulations,
    a much smaller and hence computationally less expensive simulation box is used (see \Cref{fig:setup}b).
    This thin cut-out of the total cell has a voxel count of $18\times3\times3$ and may be regarded as a pseudo-1D setup,
    mainly used for parameter analysis or long-term cycling.
    In both cases,
    the solid diffusion in the spherical active material particles in the \ce{Ni} electrode (cf. \Cref{eq:radial_solid_diffusion} and \Cref{fig:solid_diffusion}) is discretised by five equidistant grid points,
    while the simulation domains possess a non-equidistant grid to reduce computational costs (see Section 2 of the ESI\dag).
    Initial conditions for e.g. the composition of the electrolyte solution or the initial volume ratio of \ce{Zn} to \ce{ZnO} are stated in Section 3 and 4 of the ESI\dag.
    The same applies for the general boundary conditions for the applied current or voltage (Section 5 of the ESI\dag),
    which are based on the cycling protocols introduced in \secref{subsec:theory_experiments}{the section on experimental procedures}.
    The pseudo-1D setup possesses periodicity in z-direction.
    As this setup does not include the tab connections,
    here the current and voltage boundary conditions are applied to the back of the respective current collectors.
    For this purpose,
    an artificial and very thin current collector is added to the backside of the \ce{Ni} electrode (not depicted in \Cref{fig:setup}b,
    as the actual \ce{Ni} electrode uses a \ce{Ni} mesh as a current collector).
    \par
    We estimated most material constants such as ion diffusion coefficients or conductivities from literature as described and given in the ESI\dag,
    but some parameters lay in a large range of values due to composition, uncertain additives or other influences,
    and hence have to be assumed by an elaborated guess.
    The various standard reduction potentials $\Delta\phi^{\circ}$ of the electrochemical reactions for example have to be adapted to the conditions in the battery cell,
    but as they follow more specific considerations such as the influence of electrolyte concentration or known additives,
    they have relatively fixed values discussed in Section 9 of the ESI\dag.
    Other parameters with a certain degree of uncertainty are the reaction rate constants $k^i$,
    the proton diffusion coefficients $D^{\ce{NiO(OH)}}$/$D^{\ce{Ni(OH)2}}$,
    interaction coefficient $\gamma$ and voltage hysteresis $\Delta U^\text{hyst.}$ of the \ce{Ni} active material,
    and the Bruggeman coefficient of the \ce{ZnO} film around the zinc particles,
    $\beta^{\ce{ZnO}\text{-L}}$.
    A suitable and meaningful set of values for these parameters is listed in \Cref{tbl:fitted_parameters}.
    \begin{table*}
      \begin{center}
        \caption{Estimated parameters of the electrochemical \ce{Ni}/\ce{Zn} cell model.}
        \label{tbl:fitted_parameters}
        \begin{tabular*}{\textwidth}{@{\extracolsep{\fill}}llllll}
          \toprule
            $\bm{k^\textbf{I}}$ & $\bm{k^\textbf{II}}$ & $\bm{k^\textbf{IV}}$ & $\bm{k^\textbf{V}}$ & $\bm{k^\textbf{VI}}$  & $\bm{k^\textbf{VII}}$  \\
            (\unit{\mol\per\square\centi\metre\per\second}) & (\unit{\mol\per\square\centi\metre\per\second}) & (\unit{\mol\per\square\centi\metre\per\second}) &(\unit{\mol\per\square\centi\metre\per\second}) & (\unit{\per\second}) & (\unit{\per\second}) \\
          \midrule
          \num{2.5e-10} & \num{1.0e-7} & \num{2.5e-21} & \num{1.5e-18} & \num{5.0e-3} & \num{3.0e+2} \\
          \bottomrule
        \end{tabular*}
        \begin{tabular*}{\textwidth}{@{\extracolsep{\fill}}llllll}
          \toprule
            $\bm{k^\textbf{VIII}}$ & $\bm{D^{\textbf{\ce{NiO(OH)}}}}$ & $\bm{D^{\textbf{\ce{Ni(OH)2}}}}$ &  $\bm{\beta^{\textbf{\ce{ZnO}\text{-L}}}}$ & $\bm{\gamma}$ & $\bm{\Delta   U^\text{\textbf{hyst.}}}$\\
            (\unit{\per\second}) & (\unit{\square\centi\metre\per\second}) & (\unit{\square\centi\metre\per\second}) & (---) & (---) & (\unit{\volt})\\
          \midrule
          \num{2.5e-1} & \num{9.5e-11} & \num{1.0e-12} & \num{1.75} & \num{-8} & \num{0.1325} \\
          \bottomrule
        \end{tabular*}
	  \end{center}
    \end{table*}
    These parameters were chosen to catch typical characteristics of \ce{Ni}/\ce{Zn} cells,
    as they are depicted in the \secref{sec:results_and_discussion}{result section},
    see \Cref{fig:envelop_curves_and_fit}a and \Cref{fig:longterm}a.
    This includes e.g. the voltage plateau at the end of charge,
    C-rate dependence of the charge/discharge branch,
    infliction points of the cell voltage and the Coulombic efficiency (CE) of the regular cycles.
    Generally,
    the reproduction of the regular cycles is prioritised over formation cycles,
    as they make up the main part of the cycle life.
    \par
    The model described above and in detail in Sections 6 to 11 of the ESI\dag is implemented in the software package \textsc{BEST} (Battery and Electrochemistry Simulation Tool),\cite{BEST}
    using the finite volume method (FVM).
    The software relies on the software package \textsc{SAMG}\cite{SAMG} to solve the non-linear system of equations and allows for multiprocess execution using \textsc{OpenMP}.
    \par
    For the 3D visualisation of simulation results the visualisation software \textsc{VisIt}\cite{Childs2012,VisIt333} is used.
  \subsection*{Experimental procedures}
  \label{subsec:theory_experiments}
    We conduct the following experiments with a \SI{8}{\amperehour} \ce{Ni}/\ce{Zn} cell prototype (\textsc{SunErgy}).
    \paragraph{Cycling data}
    \label{para:method_cycling}
    Several cycling tests with prototypes of slightly varying parameters,
    e.g. \ce{Zn} particle size,
    have been performed.
    All prototypes are very similar to the prototype cell described in the ESI\dag.
    \par
    During initial cycling of a pristine cell,
    the first three cycles employ a specialised cycling protocol to form and condition the \ce{Zn} electrode (formation cycles).
    This deliberate, slow-rate charge/discharge procedure is essential for establishing the initial pore structure of the \ce{Zn} electrode through the conversion of \ce{ZnO} to metallic \ce{Zn},
    which exhibits a lower molar volume.
    Furthermore,
    this process enhances the conductive network of the metallic zinc.
    The formation protocol follows conventional practice:
    charging at low current density until reaching the characteristic voltage plateau of \ce{Ni}/\ce{Zn} cells,
    followed by a likewise slow discharge.
    Upon completion of the formation cycles,
    regular cycling commences, representing the intended operational conditions for this prototype.
    The standard cycling protocol initiates with charging at a $1C$ rate until the cell voltage reaches \SI{1.94}{\volt}.
    This is followed by constant voltage (CV) charging maintained until either the projected capacity of $Q^{\text{max}} = \SI{8}{\amperehour}$ has been transferred
    or the current decreases below \nicefrac{C}{20} as a safety cutoff.
    The discharge phase proceeds at a $1C$ rate until the cell voltage falls below \SI{1.0}{\volt}.
    \par
    All cycling data has been recorded with a BST8-12 battery analyser from MTI (\SI{5}{\milli\ampere} – \SI{12}{\ampere} up to \SI{5}{\volt}) following the cycling protocol described above.
    The data points have been recorded every \SI{120}{\second}.
    Additions of few grams of deionised water may have been made every few hundreds of cycles,
    if the level of electrolyte decreased because of the water splitting due to \cref{rxn:side_reactions_I,rxn:side_reactions_II}.
    The combination of cell composition and cycling protocol allows for more than 3800 cycles.
    \par
    We present the cycling data in the form of envelope curves (see \Cref{fig:envelop_curves_and_fit}a,b),
    which are in the case of the formation cycle a selection of the third formation cycle of different realisations of the prototype.
    For the regular cycle,
    the envelope curve comprises a selection of cycles between cycle number 49 and 101 of one realisation of the prototype.
    \paragraph{$\bm{\mu}$-XRF measurements}
    \label{para:method_xrf}
    For the comparison of pristine and end-of-life distribution of elemental zinc,
    another prototype cell has been cycled for 1800 cycles.
    While the formation cycles followed the same procedure as above,
    the regular cycles had an altered protocol:
    Constant current (CC) charging was performed at a $\nicefrac{C}{3}$ rate until $Q^{\text{max}} = \SI{8}{\amperehour}$ has been transferred.
    Discharging happened then again with a $1C$ rate until the cell voltage dropped below \SI{1.0}{\volt}.
    \par
    A micro X-Ray fluorescence ($\mu$-XRF) device (Brucker M4 Tornado) was employed for spatially resolved elemental analysis of the Zn electrode.
    The $\mu$-XRF images depicting the elemental zinc distribution in the electrode are shown in \Cref{fig:mu-xrf-3D-Zn}a and b.
\section*{Results and Discussion}
\label{sec:results_and_discussion}
  We investigate the cycling behaviour of the virtual \ce{Ni}/\ce{Zn} cell described in the \secref{subsec:theory_modelling}{modelling} and \secref{subsec:setup}{setup} sections,
  making direct comparisons with experimental results obtained via methods outlined in the \secref{subsec:theory_experiments}{experiment section}.
  The core cycling characteristics of our model are evaluated against experimental data,
  and long-term performance is analysed to assess model quality and to identify potential improvements for \ce{Ni}/\ce{Zn} cells.
  Results are presented and discussed systematically by topic.
  \subsection*{Charge and discharge of the cell}
  \label{subsec:discussion_behaviour}
\begin{figure}[htb]
    \centering
      \includegraphics[width=8.4cm]{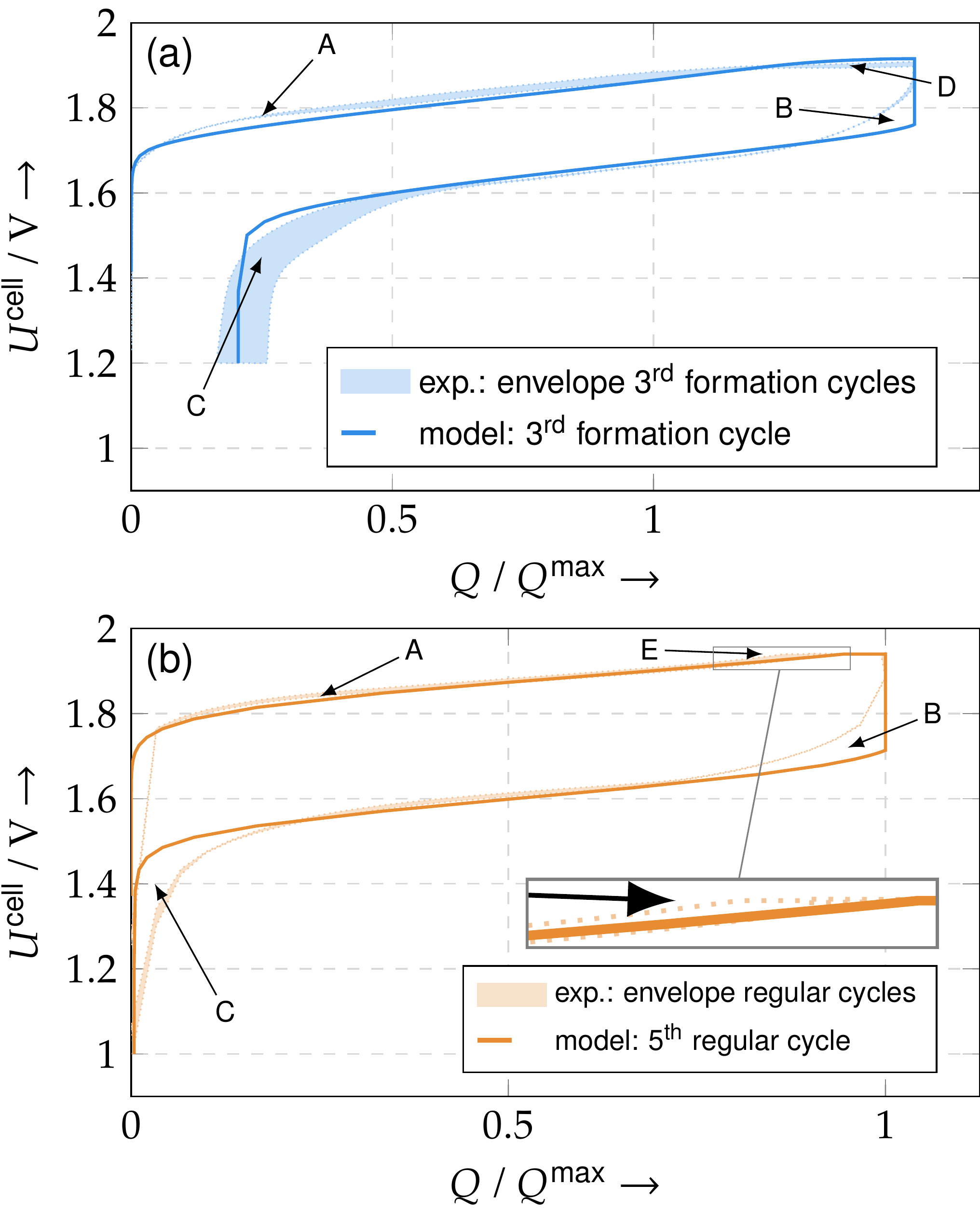}
      \caption{Charge and discharge behaviour of the prototype and virtual cell.
               (a) Experimental envelope and simulated curves of the cell voltage over transferred charge of the 3\textsuperscript{rd} formation cycle.
               (b) Experimental envelope and simulated curves of the cell voltage over transferred charge of the 5\textsuperscript{th} regular cycle.
               Segments of interest in both curves have been marked by letters A to E.
      }
      \label{fig:envelop_curves_and_fit}
    \end{figure}

    In a first step,
    we only carry out a few cycles and lay an emphasis on the behaviour of the model during a single charge/discharge cycle as well as how it relates to experimental observations.
    \paragraph{Cell voltage during one cycle}
    %
    %
    %
    %
    The cycling experiments of the slightly varied prototype cells have been collected in an envelope curve for the third formation cycle and typical regular cycles,
    see \Cref{fig:envelop_curves_and_fit}.
    Both cycle types show typical behaviour for this cell type.
    The formation cycle (blue shaded) attempts a full charge,
    which gives a steadily increasing cell voltage with a slightly convex trend.
    This increase is slowly diminished and ends in a voltage plateau soon after the projected \SI{8}{\amperehour} of transferred charge are surpassed.
    From here,
    the cell voltage does not drop abruptly onto the discharge curve but joins it smoothly,
    before it collapses at the end of discharge.
    In principle,
    this behaviour is also maintained at the higher C-rates of the regular cycles (orange shaded).
    Differences show of course in the level of the cell voltage,
    the cycling protocol related voltage plateau and the slightly more convex shape of the charging branch.
    \par
    These experimental curves have been superposed by the simulation results (solid lines) obtained as described in the \secref{subsec:setup}{simulation section}.
    We observe that for the larger part of the charge and discharge curve a close agreement between experiments and simulation is reached.
    To a certain extent,
    this is also true for parts where a deviation may be observed,
    as the same trends are visible.
    For an easier identification,
    these segments,
    where differences between model and experiment show,
    have been labelled from A to E.
    \par
    At the beginning of charge (segment A),
    experimental cycling curves show a slight rise,
    more pronounced in the formation cycle.
    The simulated voltage maintains a constant slope.
    During discharge (segment B),
    experiments exhibit a small voltage drop followed by a smooth decline forming an arc,
    equally observed in slow and fast discharging,
    i.e. formation and regular cycles,
    respectively.
    Simulations, however,
    predict this as a nearly instantaneous drop with a mild arc.
    Near the end of discharge (segment C),
    cell voltage declines less abruptly in experiments compared to simulations.
    Two unique cycle-specific segments are notable:
    the formation cycle's end-of-charge plateau (segment D),
    which the model reproduces imperfectly,
    and the regular cycle's pre-CV charging stage (segment E),
    where the cell voltage curve becomes convex (see magnification in \Cref{fig:envelop_curves_and_fit}b),
    also predicted by simulations but to a lesser extent.
    \\~\\
    %
    %
    %
    When examining the course of the simulated charge/discharge curve in \Cref{fig:envelop_curves_and_fit}a and b,
    we see a reasonable reproduction of the respective experiments for both,
    the formation and regular cycle.
    This underlines the validity of our modelling approach,
    which shows in several aspects of the curves.
    Generally,
    the combination of the models for the conversion-type \ce{Zn} and insertion-type \ce{Ni} electrodes is well-suited to depicting the slight but constant increase in cell voltage in the charging phase,
    as well as the decreasing counterpart during discharging.
    The kinetics of \cref{rxn:elchem_reactions_I,rxn:elchem_reactions_II,rxn:side_reactions_II} and their overpotentials together with a reasonably-scaled voltage offset used in our model (cf. Section 9 of the ESI\dag) describe accurately enough the height of the voltage drop when switching from charging to discharging,
    and in conjunction with the solid diffusion proton transport into the \ce{Ni} active material,
    the cell voltage breakdown at the end-of-discharge is sufficiently matched.
    Besides this,
    further issues are covered,
    which are most prominently the end-of-charge voltage plateau induced by the OER (formation cycle, blue),
    suppressing the extraction of protons from the active material and keeping the cell voltage at a level comparable to the experimental findings,
    and the behaviour when switching from CC to CV charging in the regular cycle (orange).
    There,
    the cell voltage development is close to the experiments,
    which shows especially in the continued charging operation after switching to CV indicating a suitable modelling of the continued proton extraction from the \ce{Ni} active material.
    Since all of this also happens sufficiently accurately at the different C-rates involved when charging and discharging in formation and regular cycles,
    the model is suitable for realistic simulations of a \ce{Ni}/\ce{Zn} cell.
    \par
    However,
    the simulations are not identical to the experiments,
    as shown in points A to E marked in \Cref{fig:envelop_curves_and_fit}a and b.
    These deviations are mostly rooted in the compactness of the model,
    which leaves out details,
    e.g. the influence of additives as \ce{Ca(OH)2} in the \ce{Zn} electrode (see Table S3 in the ESI\dag),
    or uses simplified approaches,
    e.g. for multi-step \cref{rxn:elchem_reactions_I,rxn:side_reactions_I,rxn:side_reactions_II}.
    The latter certainly affects the rate-dependence of the model,
    and more detailed descriptions would be possible,
    e.g. to capture the complexity of the OER.\cite{Eberle2014}
    Although these deviations are considered minor in the context of the overall functioning of the model,
    we may still obtain a better understanding of the model and relevant processes in the \ce{Ni}/\ce{Zn} cell by assessing them.
    \par
    %
    The slight voltage bump in \textit{segment A},
    which wears off in the course of charging,
    indicates an initial, temporary transport hindrance.
    This may be linked to the permeability of the \ce{ZnO} layer or the proton diffusion in the \ce{Ni} active material.
    Both have their most unfavourable configuration at this stage,
    i.e. maximum layer thickness and highest amount of diffusion-inhibiting \ce{Ni(OH)2}.
    Indeed a bump may be induced by e.g. varying the Bruggeman coefficient $\beta^{\ce{ZnO}\text{-L}}$ or the diffusion coefficient $D^{\ce{Ni(OH)2}}$.
    However,
    these effects should scale with the C-rate,
    but the bump seems to be lower in the regular ($1C$) than in the formation cycle (\nicefrac{C}{10}).
    This may rather imply a connection to the \ce{Zn} electrode formation process,
    in which large parts of the initial, transport-hindering \ce{ZnO} is converted bit by bit to \ce{Zn}.
    Other influences such as the role of \ce{Ca(OH)2} or non-wetted active surfaces due to the convection of the electrolyte solution may play an additional role.
    \par
    %
    In \textit{segment B},
    the behaviour of the cell voltage consists of two parts:
    a small instantaneous drop followed by a gradual decline.
    While the former may be attributed to the change in reaction overpotentials from charge to discharge operation,
    the latter indicates an uncovered time- or charge-dependent process.
    This process affects only that part of the voltage difference between charge and discharge branch,
    which is commonly identified as the hysteresis of the $\beta$-\ce{NiO(OH)}/$\beta$-\ce{Ni(OH)2} electrode.\cite{Ta1999,Srinivasan2000,Srinivasan2001,Albertus2008}
    The true physical reason for this hysteresis is still unknown,
    but a re-ordering or transformation process in the \ce{Ni} active particles would explain the slow drop in voltage.
    Due to the unclear origin,
    we model this hysteresis by a simple voltage offset between discharge and charge branch (cf. Section 9 of the ESI\dag),
    which explains the deviation between simulation and experiment.
    %
    \par
    %
    In \textit{segment C},
    the voltage collapse at the end of discharge happens,
    which is accurately predicted by position but not so much in shape.
    During discharging,
    \cref{rxn:elchem_reactions_II} inserts protons into the outer shell of the \ce{Ni} particles which diffuse into the active material.
    This diffusion gradually slows down as more protons are inserted,
    and at one point protons are inserted into this outer shell faster than they may be transported deeper into the particle.
    When the outer shell reaches its limit,
    the voltage collapses.
    In addition,
    the concentration overpotential of \cref{rxn:elchem_reactions_II} influences the rough voltage curve at this point.
    Thus,
    possible remedies could involve the introduction of either \ce{Ni} particles of different size to smear out the voltage drop or surface concentrations at these particles to have a stronger influence of the concentration overpotential.
    Neither of them is used here for the sake of model simplicity.
    A different possibility would involve varying the relevant reaction constant $k^\textrm{II}$,
    which may smooth this segment as discussed in Section 12.1 of the ESI\dag.
    Due to its strong influence on an overall consistent cell behaviour,
    this possibility was ruled out.
    \par
    %
    The behaviours observed in segments D and E represent distinct electrochemical processes that are interlinked.
    We analyse each segment separately before examining their mutual influences.
    \par
    At \textit{segment D},
    during the formation cycle,
    a voltage plateau is reached due to a dominating OER at the end of charge.
    The \ce{O2} formation is in competition with the proton insertion reaction,
    which shifts into a less favourable overpotential region when most protons have been extracted from the active material.
    The slightly underestimated OER resulting in a higher voltage plateau compared to the experimental curves could easily be corrected by modifying its reaction constant $k^\textrm{V}$,
    which also controls the onset of the very same plateau.
    \par
    \textit{Segment E},
    see the magnification in \Cref{fig:envelop_curves_and_fit}b,
    points out the region before the cycling protocol switches from CC to CV charging,
    which happens upon reaching a cell voltage of $U^\text{cell} = \SI{1.94}{\volt}$.
    Due to its convex voltage curve,
    the experiments attain this limit earlier than the simulation,
    which remains relatively flat.
    This onset of convexity indicates that similar to what has been discussed for segment C,
    the \ce{Ni} particles' outer shells are nearly depleted of protons through the extraction \cref{rxn:elchem_reactions_II} resulting in a raise of the SOC-dependent OCP of the very same reaction.
    At the same time,
    the OER seems not to be strong enough yet to suppress this convexity,
    which is beneficial for the Coulombic efficiency of the cell.
    To bring the simulation closer to the experimental observation,
    the solid diffusion coefficients $D^{\ce{NiO(OH)}}$ and $D^{\ce{Ni(OH)2}}$ have to be reduced,
    so that the extraction limit is reached earlier and a convexity is established.
    However,
    this may also influence the overall steepness of the charging branch.
    \par
    After the mechanisms of the two segments have been considered individually,
    their interaction and its consequence for the \ce{Ni}/\ce{Zn} cell are discussed.
    The use of a stronger OER to adjust the voltage plateau (segment D) inevitably shifts its onset,
    which may already flatten the voltage curve in segment E,
    restraining the desired convex behaviour.
    This results in a reduced Coulombic efficiency.
    Conversely,
    changing the solid diffusion constants to tune the behaviour in segment E increases the cell voltage earlier on,
    which may translate to overpotentials at the \ce{Ni} electrode that favour the OER too soon,
    leading again to a lower Coulombic efficiency.
    This may be counteracted by attenuating the OER,
    but with the negative effect of a voltage plateau in segment D,
    which deviates from the experimental observation.
    The approximation of these two processes in our model is hence a compromise,
    where both segments have a small deviation from the experimental data,
    while the Coulombic efficiency reaches realistic values as discussed in the next section.
    \par
\begin{figure}[htbp]
      \centering
        \includegraphics[width=8.4cm]{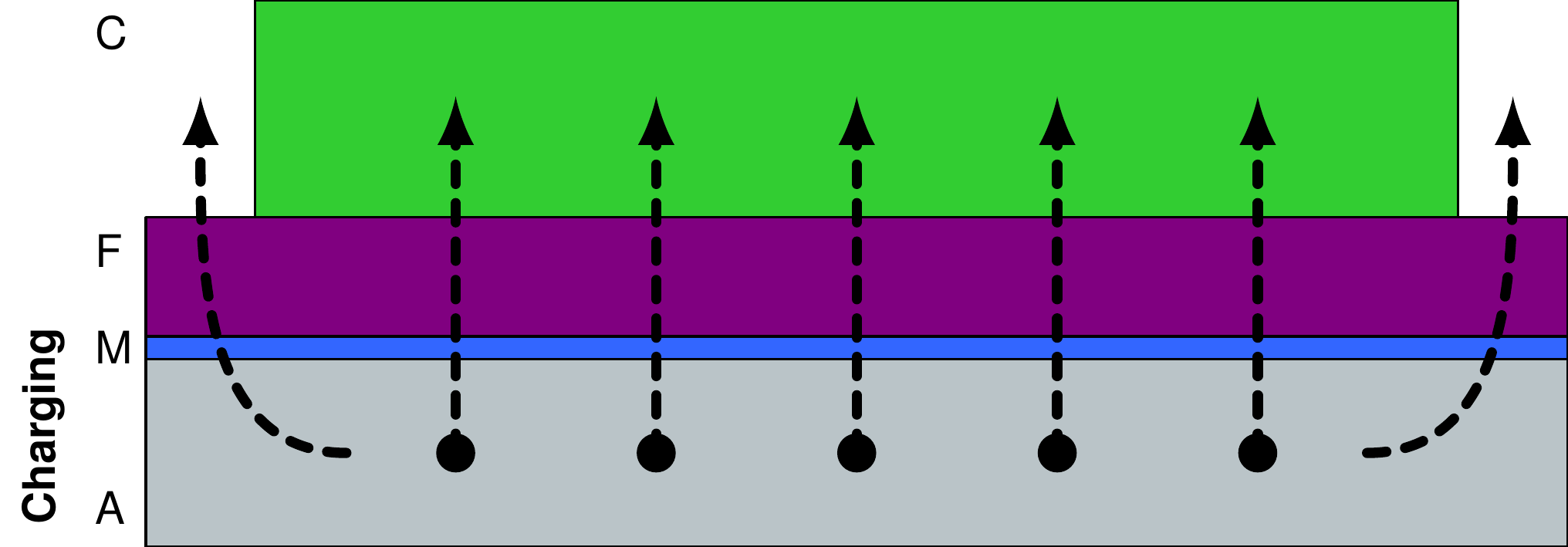}
        \caption{Schematic movement of the electrolyte solution in the virtual cell (dashed arrow) during charge.
                 The liquid is forced out of the \ce{Zn} electrode.
                 When discharging, the direction is reversed.
                 The different regions mark the \ce{Zn} electrode/anode (A),
                 the membrane (M),
                 the felt (F)
                 and the \ce{Ni} electrode/cathode (C).}
        \label{fig:convection}
    \end{figure}

    A secondary consequence of this interplay is the placement of the cycled \SI{8}{\amperehour} capacity window within the total state of charge (SOC) range during regular cycling (discussed in the following section, see \cref{fig:longterm}c).
    The faster the solid diffusion is assumed in our model,
    the lower SOCs are possible at the end of discharge,
    which pushes the cycled capacity window away from the beginning of the OER.
    However,
    for realistic values of the solid diffusion coefficients,
    the upper end of this window is at the onset of the OER,
    which agrees with experimental observations and thus speaks in favour of a good approximation by the model.
    As a result,
    this further narrows down the scope of the involved parameters $k^\textrm{V}$,
    $D^{\ce{NiO(OH)}}$ and $D^{\ce{Ni(OH)2}}$.
    More profound approaches for a better approximation of the experimental cell voltage curve may therefore be,
    as before,
    a more complex modelling in the form of e.g. a multi-step reaction for the OER or the use of a particle size distribution to disperse the behaviour of the solid diffusion.
    \paragraph{Convection during one cycle}
    %
    %
    When examining the cycling behaviour of the virtual cell,
    another important detail is the convection of the electrolyte solution in it.
    Its observed movement during charge has been schematically illustrated in \Cref{fig:convection}.
    We see that liquid moves out of the \ce{Zn} electrode (A) towards the membrane (M), felt (F) and ultimately the \ce{Ni} electrode (C) and the regions next to it.
    This happens strongly through central pathways,
    but also involves movement through the outer parts of the cell components.
    When discharging,
    this process is reversed,
    and liquid is sucked back into the \ce{Zn} electrode.
    \par
    %
    %
    The model-based convection behaviour shown in \Cref{fig:convection} suggests a counter-intuitive flow of electrolyte solution during cycling.
    By just reasoning on the basis of the \ce{Zn}/\ce{ZnO} conversion,
    the charge operation should lead to a significantly enlarged pore space due to the \SI{40}{\percent} smaller molar volume of \ce{Zn},
    sucking the electrolyte solution into the \ce{Zn} electrode as e.g. assumed by Choi and co-workers\cite{Choi1976,Choi1979}.
    But when accounting for all species consumed and created in \Cref{rxn:elchem_reactions_I,rxn:chem_reaction},
    based on our parameterisation,
    the volume of the electrolyte solution expands more than the pore space (approximately in a ratio of 4 to 1).
    This trend also applies to the single steps of the overall dissolution-precipitation process.
    Hence,
    a counter-intuitive convection scheme is found,
    and in its consistency along the whole reaction chain it may indicate a prominent role in the cell.
    This reasoning has already been made by Einerhand \textit{et~al.}\cite{Einerhand1991a,Einerhand1991b} on an experimental basis,
    and has been described under the term \textit{density gradient model} in their work.
    They note that the convection pattern plays an important role in zincate redistribution and hence is partly responsible for the zinc shape change, 
    and also rule out other ideas as the (sole) influence of the membrane or concentration cells given by other authors\cite{McBreen1972,McBreen1978,Gunther1987}.
    Additionally,
    as a supporting evidence,
    our cycling experiments showed as well a slightly increasing electrolyte solution level during discharge.
    \par
    For a more precise analysis,
    however,
    other influences such as the slight volume change of the \ce{Ni} electrode when \ce{H+} insertion/extraction is happening
    and the headspace volume should be taken into account.
    %
    %
    %
  %
  %
  %
  \subsection*{Long-term cycling of the cell}
  \label{subsec:discussion_lt_effects}
\begin{figure}[htb]
      \centering
        \includegraphics[width=8.4cm]{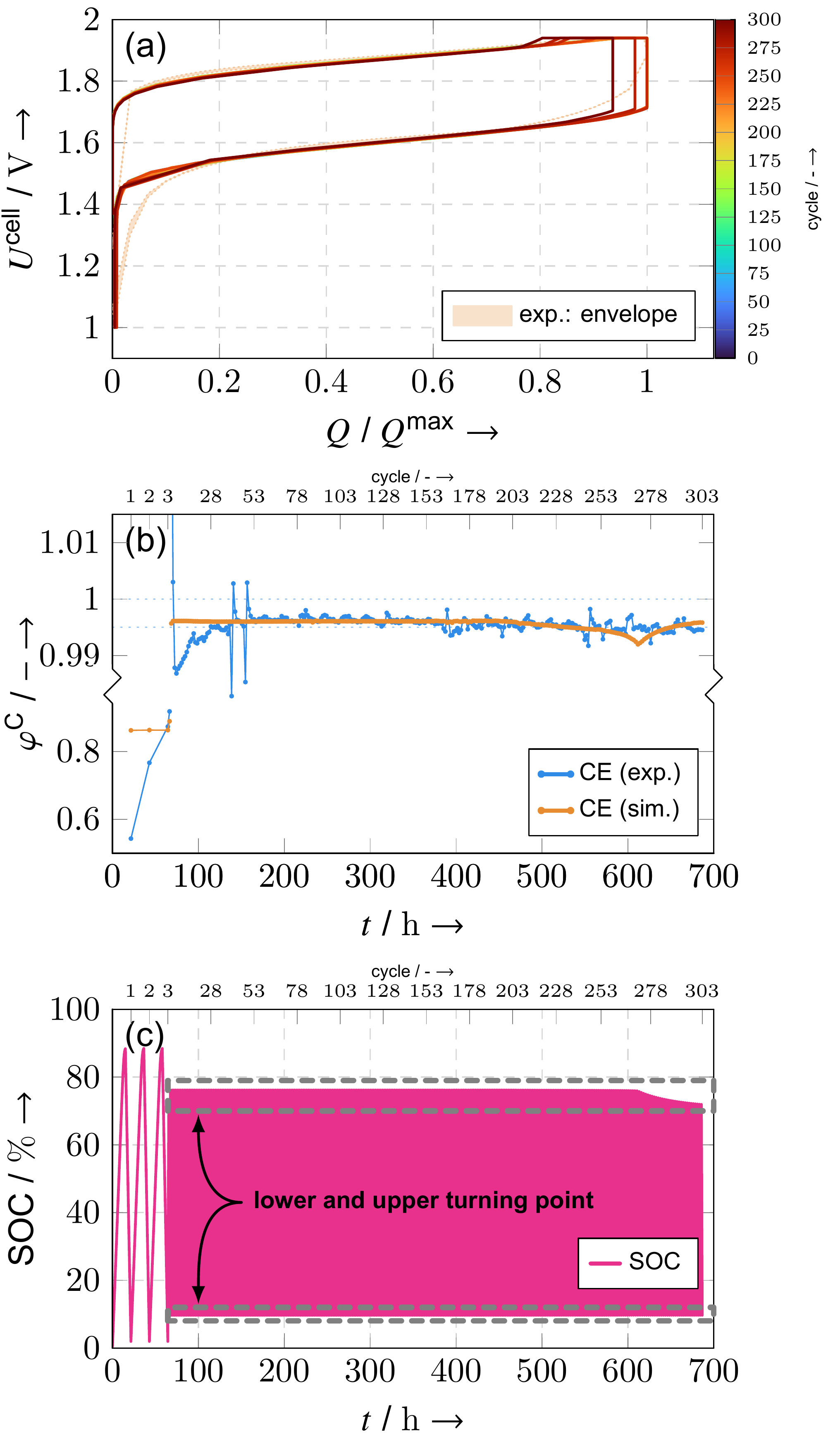}
        \caption{Results and comparison of long-term cycling.
                 (a) Cell voltage over transferred charge: regular cycles compared to experimental envelope with simulation failure starting at regular cycle 269 (equals total cycle 272).
                 (b) The development of the Coulombic efficiency (CE) and
                 (c) of the SOC over time.
                 The CE is always depicted at the end of a charge/discharge cycle and compared to the CE of a cycling experiment.
                 The CE y-axis' upper part is differently scaled to better show details around $\varphi^\text{C} = 1$.
                }
        \label{fig:longterm}
    \end{figure}

\begin{figure*}[htbp]
      \centering
        \includegraphics[width=17.4cm]{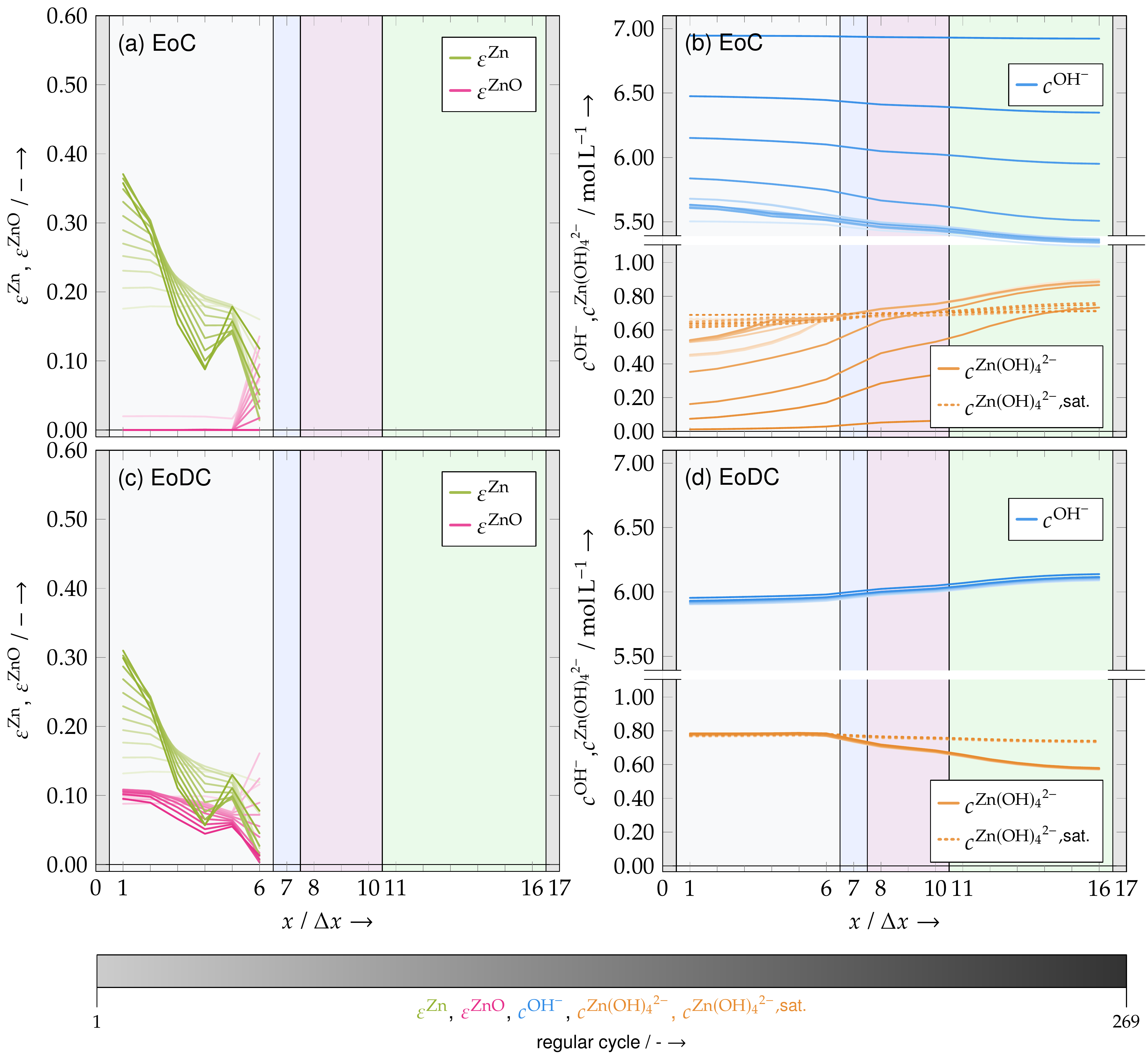}
        \caption{Evolution of volume fractions and concentrations along the x-direction of the cell.
                 Regular cycles 1, 25, 50, 75, 100, 125, 150, 175, 200, 225, 250 and 269 are depicted with lines of increasing colour intensity.
                 (a) and (c) show the volume fractions of \ce{Zn} and \ce{ZnO} at the end of charge and end of discharge, respectively.
                 Similarly, (b) and (d) show the concentration profiles of \ce{OH-}, \ce{Zn(OH)4^{2-}} and its saturation concentration at the end of charge and end of discharge, respectively.
                 The background colours indicate the compartment of the cell according to \Cref{fig:setup}.
                }
        \label{fig:profiles}
    \end{figure*}

    %
    %
    %
    For this section,
    the previously presented simulation is continued until the virtual cell starts to fail.
    This happens after approximately $269$ regular cycles when the capacity starts to decline,
    see \cref{fig:longterm}.
    The corresponding evolution of the internal configuration of the battery cell is shown in \cref{fig:profiles}.
    Furthermore,
    a 3D simulation is done for an evaluation of the zinc shape change process. 
    \paragraph{General cell behaviour}
    %
    In a first step we compare the results of our simulation to the characteristic quantities of the prototype cell,
    namely the cell voltage $U^\text{cell}$ and the Coulombic efficiency (CE) $\varphi^\text{C}$.
    This is accompanied by a glimpse on the cycled SOC window.
    \par
    In \cref{fig:longterm}a,
    we see cell voltage curves during regular charge/discharge cycles over transferred charge normalised to the cell's \SI{8}{\amperehour} capacity.
    In the background,
    as before,
    we see the envelope of several cycles of the experiment,
    which lie in the range of cycle numbers 49 to 101 (cf. \secref{subsec:theory_experiments}{section on experimental procedures}).
    On top of this,
    the simulation's cell voltage curves are drawn,
    from dark blue to dark red in ascending cycle numbers.
    Since most of the visible colours are dark red,
    this means that the cell voltage curve does not change a lot during the cycling simulation,
    except at the very end of cycle life,
    when the point at which the cycling protocol switches from CC to CV charging shifts to the left and later the charging finishes prematurely.
    \par
    \Cref{fig:longterm}b shows the Coulombic efficiency of our simulations (orange) and of the experiments performed (blue) over time during cycling.
    The two distinguishable time segments are the formation cycles 1 to 3 and the subsequent regular cycles until the start of cell failure.
    During the formation cycles,
    the efficiency in the experiments increases from approximately $0.55$ to $0.87$,
    while the simulation shows a steady value of $0.85$.
    When switching to regular cycles,
    the overall behaviour of experiment and simulation is similar:
    The Coulombic efficiency stabilises in a range between $0.995$ and close to $1.0$ (indicated by the two blue dotted lines).
    The simulation result fluctuates in a barely visible, much narrower range.
    Differences are noticeable in two points.
    Firstly,
    at the beginning of the regular cycles,
    the experiments show an initial spike followed by a short dip,
    whereas the simulation directly attains stable values.
    Secondly,
    at the end of the cycle life of the virtual cell,
    its CE slowly drops until a visible minimum at regular cycle $269$ before rising again afterwards.
    In contrast,
    at this point,
    the prototype cell is still far from cell failure and shows therefore no unusual features.
    \par
    The last panel,
    \cref{fig:longterm}c,
    depicts the development of the simulated cell's SOC over time.
    Again,
    the two segments of formation and regular cycles are clearly separable.
    The SOC ranges in the formation cycles from nearly \SI{0}{\percent} to approximately \SI{90}{\percent} and in the regular cycles in a window between approximately \SI{10}{\percent} and \SI{76}{\percent}.
    After regular cycle $269$,
    the SOC and hence the available capacity starts to drop visibly.
    Generally we see that the lower and upper turning point of the SOC remains very stable throughout cycling.
    \\~\\
    The battery model also shows in long-term cycling simulation a behaviour which approximates the experiments closely.
    Like these,
    it predicts stable cycles (Figure \Cref{fig:longterm}a) characterised by high Coulombic efficiency (Figure \Cref{fig:longterm}b) and a constant SOC window (Figure \Cref{fig:longterm}c) throughout cycle life.
    Even small but barely visible fluctuations in both CE and SOC are reproduced.
    \par
    %
    The two major deviations from the experimental CE are readily explained.
    During formation and initial regular cycles,
    the experimental data shows behaviour not captured by the model,
    as \ce{Zn} electrode formation processes are not covered.
    However,
    the model successfully reproduces the third formation cycle, which was used for parameter adjustment.
    Furthermore,
    the initial differences in the regular cycles are mostly due to the experiments using two other cycling protocols before switching to the one used by the simulation.
    At end-of-life,
    the virtual battery exhibits a complex behaviour.
    The internal configuration has changed to such an extent,
    i.e. zincate shortage acting on concentration overpotentials (see next paragraph),
    that the voltage limit to switch from CC to CV charging is reached earlier on (see \Cref{fig:longterm}a),
    and hence an extended period of high voltage promotes oxygen formation resulting in a phase of steadily declining Coulombic efficiency.
    At the minimum of CE,
    regular cycle $269$,
    the CV period has developed a length which lets the charging current drop below the limit defined in the cycling protocol (cf. \secref{para:method_cycling}{the experimental section}) before reaching \SI{8}{\amperehour} of transferred charge.
    In consequence,
    the SOC window cannot be maintained - the dischargeable capacity starts to diminish.
    Due to this now dominant limit for end-of-charge,
    the CV period shortens again which leads to less \ce{O2} formation and thus the CE rises again while the SOC window continues to shrink.
    At the same time,
    the again increasing CE starts to get limited by the HER,
    which slowly gains weight due to the aforementioned changes in internal configuration.
    Therefore,
    the already mentioned regular cycle $269$ poses here as an indicator of maximum cycle life.
    This is little compared to the more than $3800$ cycles reported for the experiment,
    but not unusual as our model and its parameterisation does not cover all aspects of the prototype cell.
    \par
    Nevertheless,
    the virtual cell continues to cycle until regular cycle $395$,
    in which around \SI{85}{\percent} of the initially utilised capacity is still accessed and the cell fails due to continued zinc redistribution and follow-up transport restrictions.
    %
    %
    %
    \paragraph{Internal configuration}
    %
    Beyond this rather external characterisation of the cell behaviour,
    the simulation enables us as well to examine the internal dynamics that drive the observed behaviour.
    For this purpose we analyse the evolution of the volume fractions of \ce{Zn} and \ce{ZnO} and concentration profiles of \ce{OH-} and \ce{Zn(OH)4^{2-}},
    depicted for end of charge (see \cref{fig:profiles}a,b) and end of discharge (see \cref{fig:profiles}c,d),
    respectively.
    \par
    Initially,
    \ce{Zn} and \ce{ZnO} volume fractions (\cref{fig:profiles}a) show nearly equal distribution at end of charge,
    with some variation near the membrane.
    However,
    as cycling progresses,
    \ce{Zn} increasingly accumulates toward the current collector while \ce{ZnO} quickly depletes throughout the electrode,
    last disappearing near the membrane.
    In advanced cycle stages,
    \ce{ZnO} is completely consumed across the entire electrode at end of charge while \ce{Zn} develops a small peak due to spatially different growth rates.
    \par
    The concentrations in \cref{fig:profiles}b show a certain stability up to the first 150 to 175 regular cycles,
    with the profiles of both,
    $c^{\ce{OH-}}$ and $c^{\ce{Zn(OH)4^{2-}}}$,
    remaining in a narrow range.
    For the zincate concentration,
    for some time a relatively flat profile is observed from the membrane halfway into the electrode.
    Later,
    the profiles start to deviate from this behaviour with $c^{\ce{OH-}}$ increasing and $c^{\ce{Zn(OH)4^{2-}}}$ decreasing substantially.
    Near end-of-life,
    the zincate concentration drops to near zero throughout the cell,
    while its saturation concentration remains relatively stable.
    Generally,
    the spatial distribution shows \ce{OH-} decreasing from negative to positive electrode,
    with \ce{Zn(OH)4^{2-}} exhibiting the inverse behaviour,
    including that of its saturation concentration.
    \par
    At the end of discharge,
    the trend for and the shape of the volume fraction of \ce{Zn} is similar to that at the end of charge but with lower absolute values due to zinc dissolution during discharge (\cref{fig:profiles}c).
    For \ce{ZnO},
    the volume fraction has the tendency to be higher close to the membrane at the beginning,
    but progressively shifts toward the current collector over cycling.
    \par
    In contrast to the changes observed at end of charge,
    concentration profiles at end of discharge (\cref{fig:profiles}d) remain relatively stable with increasing cycle number.
    For \ce{OH-},
    the profile increases constantly from the \ce{Zn} to the \ce{Ni} electrode,
    while zincate decreases between these two electrodes.
    Its saturation concentration shows only minimal decline.
    \\~\\
    The general observations show once more that the model captures the essence of the functioning of the battery.
    When e.g. bringing a charging process to mind,
    roughly corresponding to the differences of volume fractions and concentrations between EoC and EoDC in \cref{fig:profiles},
    the following consistent picture is drawn:
    \ce{Zn} increases in the anode through \cref{rxn:elchem_reactions_I} consuming \ce{Zn(OH)4^{2-}}.
    This causes the zincate concentration in the \ce{Zn} electrode to drop below its saturation limit,
    which results then in the dissolution of \ce{ZnO} via \cref{rxn:chem_reaction}.
    In combination,
    these two reactions release two hydroxide ions reflecting in an increase of \ce{OH-} concentration in this compartment.
    At the other electrode,
    extraction of protons from the nickel active material via \cref{rxn:elchem_reactions_II} consumes \ce{OH-},
    explaining the drop of its concentration in the cathode.
    \par
    Furthermore,
    the observations reveal critical insights into the cell functioning and failure mechanism.
    Despite the virtual cell cycling in a relatively stable manner until the very end (cf. \cref{fig:longterm}a),
    examining the changes in \ce{ZnO} volume fraction (\cref{fig:profiles}a) and \ce{Zn(OH)4^{2-}} concentration (\cref{fig:profiles}b) shows that progressive depletion of \ce{ZnO} and subsequent zincate shortage in the electrolyte solution causes an increased concentration overpotential for \Cref{rxn:elchem_reactions_I}.
    This leads to the previously observed earlier onset of the \SI{1.94}{\volt} plateau,
    ultimately initiating cell death.
    \par
    Before cell death,
    at advanced cycling stages,
    redistribution of \ce{Zn} and \ce{ZnO} fuels this problem additionally.
    The accumulation of these towards the current collector,
    which is in agreement to experiments,\cite{Caldeira2017}
    reduces pore space and progressively restricts zincate supply towards the rear parts of the negative electrode during charging,
    amplifying aforementioned mass transport limitations.
    It has been found that the observed bump in the \ce{Zn} volume fraction depends on the cycling conditions,
    whereby a longer charging period,
    i.e. a higher transferred charge,
    prevents formation.
    Furthermore,
    lowering $\beta^{\ce{ZnO}\text{-L}}$ and hence reducing the transport hindrance of the \ce{ZnO} layer has similar effects.
    \par
    This redistribution of \ce{Zn} and \ce{ZnO} is already initiated at the beginning of cycling,
    as shown by the initial dip and peak of these two volume fractions at the membrane.
    Due to transport of zincate to the negative electrode during charging,
    the \ce{Zn(OH)4^{2-}} concentration is the highest close to the membrane,
    see \cref{fig:profiles}b.
    This has two effects:
    First,
    \ce{ZnO} is not dissolved as strongly as in the rest of the anode because the zincate concentration does not drop too much below its saturation concentration.
    This has the side effect that the \ce{ZnO} layer around the \ce{Zn} particles stays thicker in this region,
    restricting the transport of \ce{Zn(OH)4^{2-}} and \ce{OH-} to the particle surfaces.
    Second,
    together with the higher bulk zincate concentration,
    this slows down the zinc deposition during charging close to the membrane.
    This inhomogeneity and reciprocation of local reaction rates and volume fractions is then perpetuated spatially and temporally.
    \par
    The diminishing zincate concentration in later stages also explains the previously observed changes in Coulombic efficiency and SOC near end-of-life.
    As zincate depletes,
    hydroxide concentration rises correspondingly,
    creating two competing effects.
    First,
    this concentration shift alters the overpotentials for both insertion and oxygen evolution reactions,
    favouring the former due to its single-step mechanism that consumes only one hydroxide molecule compared to the more complex oxygen evolution pathway.
    Consequently,
    the insertion reaction becomes increasingly dominant over oxygen evolution.
    Second,
    low zincate concentration triggers the aforementioned earlier onset of the CV charging phase through increasing the concentration overpotential of \cref{rxn:elchem_reactions_I},
    extending the period of accelerated \ce{O2} production.
    The latter effect overcompensates the former,
    explaining the intermediate CE decline observed in \cref{fig:longterm}b. 
    With continued cycling, 
    the CV charging phase reaches an extent such that at its end the current drops below the secondary limit defined in the cycling protocol,
    terminating the charging phase before having transferred \SI{8}{\amperehour}.
    This creates the local minimum of CE visible towards end of cycling.
    As zincate scarcity increases,
    the onset of the CV phase occurs progressively earlier
    which causes the charge termination to depart from the desired \SI{8}{\amperehour}.
    This reduces dischargeable capacity,
    as reflected in SOC development.
    However,
    a lower SOC also weakens the OER,
    which in turn improves CE again.
    \par
    The failure mode observed in the simulation demonstrates that near-perfect cycling between \ce{Zn} deposition during charge and \ce{ZnO} precipitation during discharge is crucial for successful \ce{Ni}/\ce{Zn} cell operation.
    This delicate \ce{Zn}/\ce{ZnO} balance requires almost perfect back-and-forth conversion,
    which is primarily achieved and reflected through a high Coulombic efficiency.
    High efficiency enables the cell to maintain \ce{Zn} and \ce{ZnO} equilibrium by keeping oxygen evolution at minimal levels.
    Additionally,
    \ce{ZnO} recovery in each cycle is partially supported by the recombination of \ce{Zn} and dissolved \ce{O2_{(aq)}} as described by \Cref{rxn:chem_reaction_II}.
    Together,
    these mechanisms contribute to cycling stability,
    but their imperfection,
    also with regard to redistribution,
    represents a primary pathway to cell failure.
    %
    %
    %
    \paragraph{Zinc redistribution}
    %
    The analysis of \cref{fig:profiles} revealed that \ce{Zn} and \ce{ZnO} redistribution may impact cell behaviour.
    To comprehensively examine this phenomenon,
    we established a three-dimensional battery simulation following the setup in \cref{fig:setup}a. 
    \par
    %
    The simulation parameters remain identical to the previous study,
    but to account for the increased \ce{Zn} electrode in the 3D geometry,
    the specific surface area for the HER is reduced by a factor of $0.3$.
    This adjustment ensures comparable \ce{H2} production to previous pseudo-1D simulations.
    To allow for better comparison,
    we applied the same cycling protocol as for the $\mu$-XRF experiments.
    \par
    The computational expense of 3D simulations,
    driven by numerous variables and non-linearities,
    limited our study to a small number of cycles.
    We avoided 2D domains to maintain realistic convection behaviour,
    though this prevented long-term studies where shape change may become more pronounced and could strongly influence battery performance.
    \begin{figure}[htbp]
      \centering
        \includegraphics[width=8.4cm]{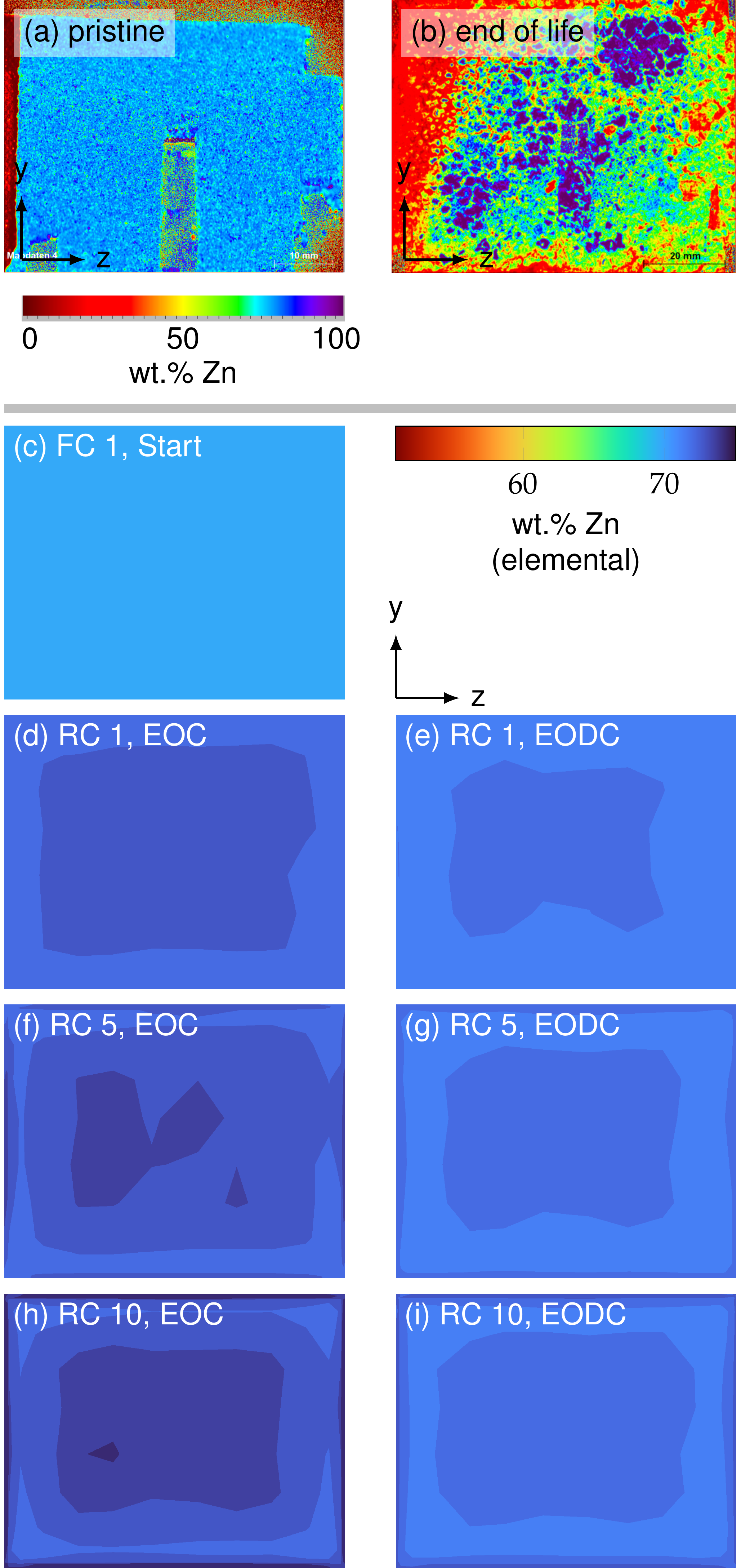}
        \caption{\ce{Zn} distribution in prototype and virtual cell.
                 (a) and (b) show $\mu$-XRF scans of the \ce{Zn} electrode before and after long-term cycling.
                 An accumulation of zinc in the centre of the electrode is visible,
                 while the outer regions deplete.
                 (c) to (i) show the backside of the virtual \ce{Zn} electrode at end of charge (EOC) and end of discharge (EODC) at different cycles.
                 Colour scheme is adapted to the experimental one,
                 but with a narrower band to visualise differences at this early stage of cycling.
                 }
        \label{fig:mu-xrf-3D-Zn}
    \end{figure}
    \par
    %
    Both,
    experiment and simulation,
    depict the elemental \ce{Zn} distribution in wt.\%,
    i.e. the zinc in metallic \ce{Zn} and \ce{ZnO}.
    In case of the experiment this is based on $\mu$-XRF measurements (for details see the \secref{subsec:theory_experiments}{experimental procedures}),
    while the simulation calculates the weight percentage assuming the electrode composition from Table S3 of the ESI\dag,
    which may deviate slightly from actual electrode values.
    \par
    \Cref{fig:mu-xrf-3D-Zn}a and b present $\mu$-XRF measurements of a prototype zinc electrode before and after extensive cycling ($1800$ cycles).
    Initially,
    elemental zinc distributes nearly homogeneously throughout the electrode.
    During cycling,
    zinc progressively migrates from outer edges and corners toward the electrode centre.
    \par
    The simulation results (\cref{fig:mu-xrf-3D-Zn}c-i) show elemental \ce{Zn} distribution viewed from the electrode backside.
    We narrowed the colour band range around initial composition values to visualise subtle distribution changes at this early cycling stage,
    while experimental images display the full range.
    The colourmap differs slightly but maintains the same red-to-blue gradient.
    \par
    Starting from homogeneous distribution in the first formation cycle (\cref{fig:mu-xrf-3D-Zn}c),
    elemental \ce{Zn} accumulates in the centre region during both formation and regular cycles.
    This accumulation persists throughout individual cycles,
    as evident when comparing end-of-charge and end-of-discharge states.
    Detailed analysis (see Figure S8 in the ESI\dag) reveals that metallic \ce{Zn} predominantly occupies the centre region while \ce{ZnO} appears off-centre.
    The simulation shows distinct behaviour along the electrode edge,
    where elemental \ce{Zn} accumulation occurs during the simulation time span - a pattern not observed in experiments.
    \\~\\
    %
    This elemental \ce{Zn} redistribution in experiment and simulation represents the well-known shape change phenomenon in \ce{Zn} electrodes\cite{McLarnon1991,Reddy2010,Mainar2016,Lu2021,Naveed2022} — a three-dimensional effect.
    Already after ten regular cycles,
    the simulation demonstrates lower elemental \ce{Zn} content at electrode edges compared to the centre in both fully charged and discharged states.
    While this difference amounts to only a few percentage points by weight at this early cycling stage,
    the model successfully captures realistic \ce{Zn} redistribution behaviour.
    \par
    The increase of elemental \ce{Zn} along electrode edges differs from $\mu$-XRF observations and represents a distinct phenomenon from classical shape change.
    This occurs because electrode edges experience heterogeneous electrode and electrolyte potentials,
    leading to different reaction rates and accelerated dendrite growth.
    To mitigate this behaviour,
    \ce{Zn} electrodes are typically oversized in y- and z-directions compared to \ce{Ni} electrodes\cite{McLarnon1991},
    moving undesired effects away from directly facing electrode regions and creating more homogeneous potentials with evenly distributed dissolution-precipitation patterns,
    thus helping to extend cycle life.
    \par
    Our model successfully demonstrates shape change capabilities in these relevant central regions.
    Comparison with experimental results (1800 cycles), \cref{fig:mu-xrf-3D-Zn}b) and literature\cite{Hendrikx1986,McLarnon1991,Einerhand1991a,Einerhand1991b} shows qualitatively similar trends.
    The uneven densification and agglomeration of elemental \ce{Zn} primarily at the electrode centre represents a well-established phenomenon.
    While ten regular cycles provide limited quantitative insight,
    subtle colour variations indicate early-stage elemental zinc unevenness that likely serves as the precursor to more pronounced zinc shape change observed in long-term cycling.
  \subsection*{Optimisation}
  \label{subsec:discussion_optimisation}
    Previous sections and analysis of parameter influences in Figure S7 of the ESI\dag reveal that the reaction rate constants of the main reactions and OER,
    amongst others,
    significantly impact the virtual battery cell's behaviour.
    This may translate similarly to (physical) \ce{Ni}/\ce{Zn} cells,
    offering room for improvements.
    Previous sections have demonstrated that the competition between OER and the insertion reaction critically determines cell efficiency,
    directly affecting the \ce{Zn}/\ce{ZnO} balance,
    which is crucial for short to medium-term cycle stability -- before longer-term effects like zinc shape change become predominant.
    Thus,
    we will focus here on ways to minimise the OER influence and hence optimise \ce{Zn}/\ce{ZnO} balance to improve the efficiency and therefore the cycle life.
    A second objective is to find out whether these efficiency gains can also be translated into higher capacity utilisation.
    \par
\begin{figure}[htbp]
      \centering
        \includegraphics[width=8.4cm]{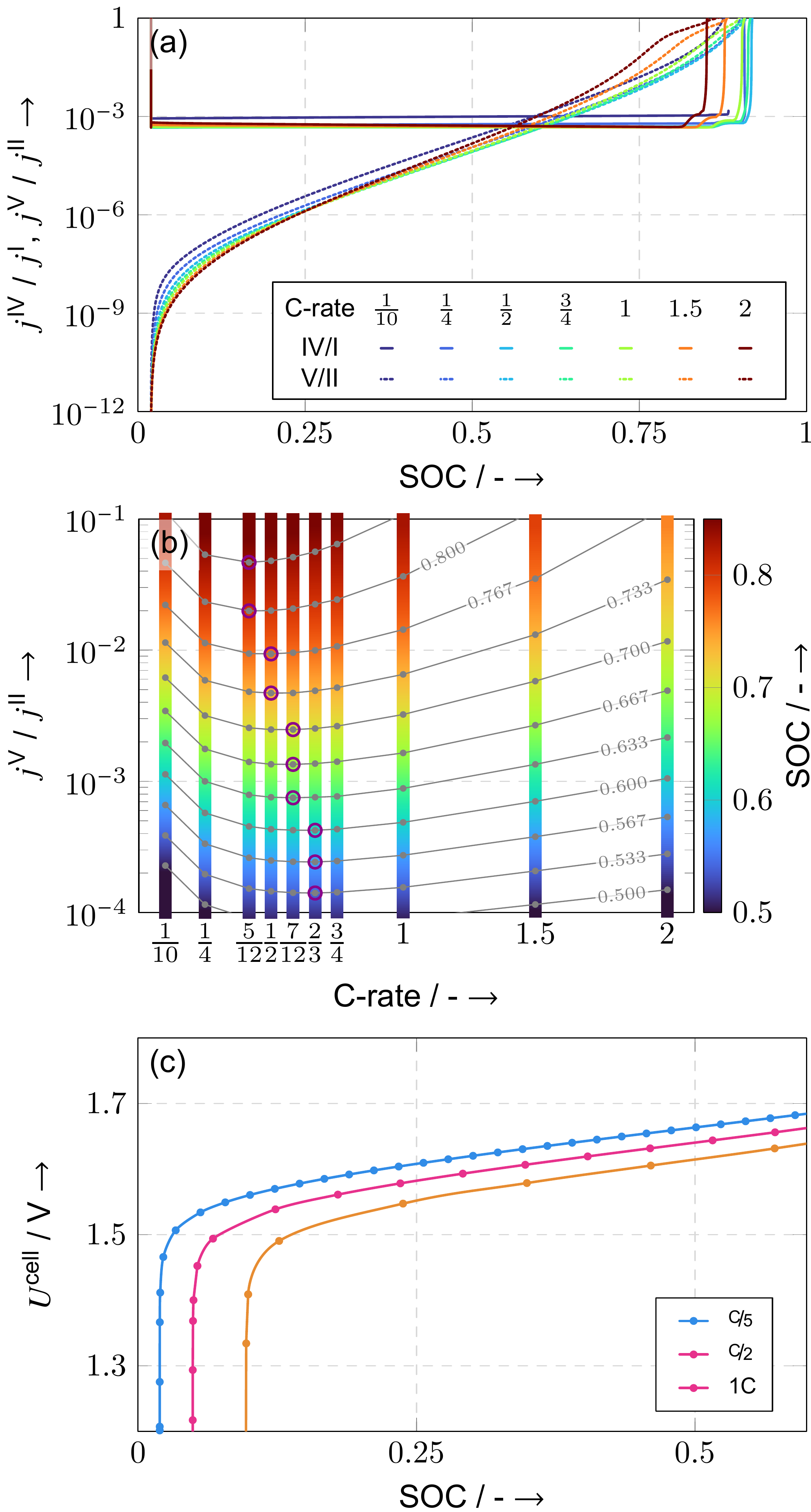}
        \caption{(a) Influence of charging rate on the competition between side and main reactions.
                     While the ratio of \Cref{rxn:side_reactions_I} to \Cref{rxn:elchem_reactions_I} (solid lines) stays approximately constant and on a low level,
                     the ratio of \Cref{rxn:side_reactions_II} to \Cref{rxn:elchem_reactions_II} (dotted lines) changes strongly.
                     Both show an optimal rate at which the influence of the side reaction is the lowest.
                 (b) Analysis of the ratio \Cref{rxn:side_reactions_II} to \Cref{rxn:elchem_reactions_II}.
                     There are optimal charging rates to minimise the influence of the OER.
                     Grey contour lines are drawn along constant SOC (linearly interpolated between points).
                     Based on the discrete data,
                     the violet circles indicate charging rates with minimal influence of the side reaction.
                 (c) Cell voltage at the end of discharge over SOC (spline through data points).
                     Lower discharging rates allow a deeper discharge.
                }
        \label{fig:reaction_competition}
    \end{figure}

    The obvious method would be to delay the onset and strength of the OER by reducing the reaction rate constant (cf. Figure S7b of the ESI\dag),
    which could correspond to the use of additives,
    for example.
    Another possibility shows,
    when plotting the ratio of the volume-averaged side reaction rate to main reaction rate for the two electrodes over the volume-averaged SOC while charging,
    the operation where the OER occurs.
    In \cref{fig:reaction_competition}a we have simulated this for several C-rates.
    The ratios for the \ce{Zn} electrode are given in solid lines,
    while dashed lines represent those for the \ce{Ni} electrode.
    %
    %
    \par
    Two observations may be made:
    Firstly,
    the ratio \textrm{IV}/\textrm{I},
    i.e. HER to zinc deposition,
    is largely invariant to the SOC.
    Only larger C-rates start to disturb this image.
    The ratio \textrm{V}/\textrm{II},
    i.e. OER to proton extraction,
    however is strongly dependent on the SOC with the OER constantly gaining weight when charging.
    Secondly,
    moderate C-rates seem to lower the ratio \textrm{V}/\textrm{II} somewhat when compared to low or high C-rates,
    which is in principle as well true for the ratio \textrm{IV}/\textrm{I},
    but to a lesser extent.
    \par
    The first observation may be explained by the fact that the overpotentials of \Cref{rxn:side_reactions_I,rxn:elchem_reactions_I} are relatively constant,
    only influenced by concentration changes which especially show at higher C-rates and SOCs.
    In contrast,
    the overpotential of extraction \cref{rxn:elchem_reactions_II} strongly depends on the SOC via the OCP curve (see Equation 21 of the ESI\dag).
    This dependence explains why the OER becomes dominant at higher SOC values,
    shifting the ratio of the two reaction rates in favour of the OER.
    The second observation however is not that trivial,
    and comprises two aspects in case of the \ce{Ni} electrode.
    For all reactions,
    the overpotentials depend on local species concentrations,
    which change proportionally to the applied charging current.
    Due to the multi-step and multi-electron nature of most reactions involved,
    concentration changes affect the overpotentials and Butler-Volmer equations differently.
    At intermediate C-rates,
    a favourable combination of concentration profiles, electrode potentials and electrolyte potentials emerges,
    benefiting the main reactions.
    At the \ce{Ni} electrode,
    the extraction reaction further influences this behaviour through the SOC-dependent OCP,
    which is determined by the local SOC in the active material particle's outer shell.
    When C-rates are high,
    this outer shell depletes more rapidly,
    making proton extraction less favourable compared to OER.
    Simultaneously,
    protons in the particle core cannot escape due to solid diffusion being slower than the reaction rate,
    causing a low SOC.
    At intermediate C-rates,
    however,
    extraction rate and solid diffusion remain sufficiently balanced,
    allowing the system to reach high SOC values before OER becomes the dominant process.
    \par
    Analysing this rate dependence further in \cref{fig:reaction_competition}b,
    the previous observations are visible in more detail.
    We see the development of the ratio of the reaction rates \textrm{V}/\textrm{II} for different C-rates along an ascending SOC.
    From the colour scheme,
    and even more from the contour lines derived from it along constant SOCs,
    we see more clearly than before that the optimal charging rate for minimum \ce{O2} production slowly shifts to lower C-rates with increasing SOC.
    Thus,
    a step-by-step reduction of the current during the charging operation may slow down the OER additionally.
    \par
    Lastly,
    in \cref{fig:reaction_competition}c,
    another optimisation approach is revealed for the discharge operation.
    There we see that slower discharge rates allow for a larger discharge depth.
    This phenomenon stems from the competition between proton diffusion speed within \ce{Ni} particles,
    as visible in Figure S7c of the ESI\dag,
    and the insertion/extraction reaction rate.
    In this sense,
    lowering the C-rate towards the end of discharge allows the solid diffusion in the particles to keep up with the reaction rate.
    This lowers the insertion rate of protons giving them more time to diffuse towards the particle centres,
    which delays reaching the proton capacity in the outer shell and hence the collapse of the cell voltage.
    Therefore this effect may be used to either further extent the cycle life by shifting the SOC window away from OER dominance or to increase the cycled capacity.
    \par
\begin{figure*}[htb]
      \centering
        \includegraphics[width=17.4cm]{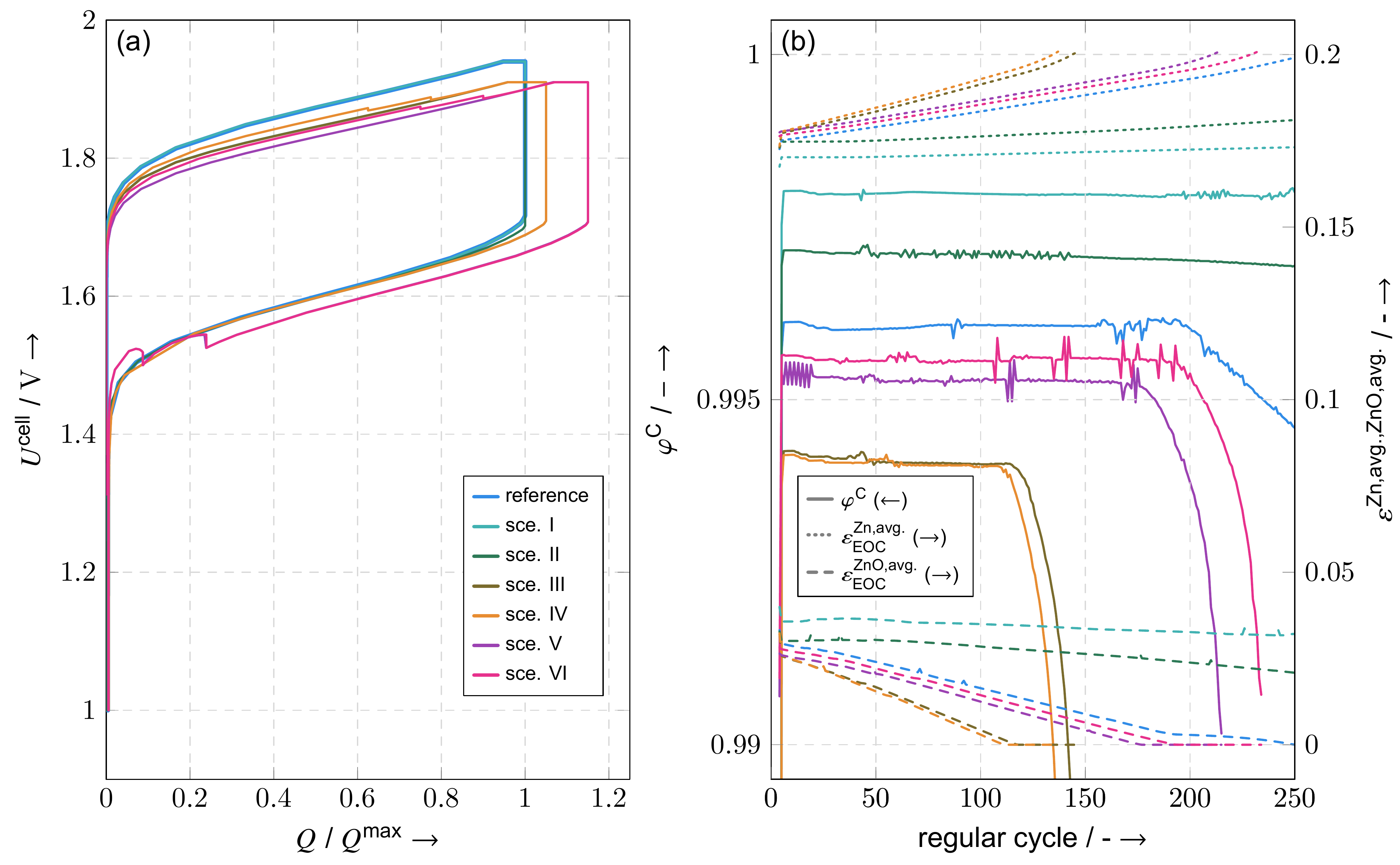}
        \caption{(a) Cell voltage during the 25\textsuperscript{th} regular cycle for the reference and derived scenarios,
                 (b) evolution of the Coulombic efficiency and volume-averaged \ce{Zn}/\ce{ZnO} volume fractions (at end of charge) over the cycle number with the same colours as in (a).
                 Curves depicted up to the cycle where capacity fading starts.
                }
        \label{fig:optimisation_oer}
    \end{figure*}

    These observations now suggest the use of tailored C-rates and charging limits to improve the efficiency, cycle life or available capacity of the cell by adapting the cycling protocol.
    This is another optimisation strategy in addition to the use of additives for OER suppression mentioned at the beginning.
    \\~\\
    The two approaches are now compared to the reference case from the previous sections in six different scenarios,
    where only the first one deals with an additive-induced reduction of the OER rate and the five remaining cover a step-wise adaption of the cycling protocol to explore the potential for cycle life and capacity improvements.
    These adapted cycling protocols are listed in \cref{tbl:protocols} and the results are shown in \cref{fig:optimisation_oer}.
    \begin{table*}
    	\begin{center}
    	\caption{Cycling protocols analysed.
    	         The green and red double arrows indicate the charging and discharging operations, respectively.
    	         Scenario \textrm{I} is the reduction of the OER rate through additives.
    	         }\label{tbl:protocols}
    		\begin{tabular}{ll}	
                \toprule		
                \textbf{Sce.}& \textbf{Operations} \\
                \midrule
                \textrm{II}  &	 \textcolor{green!50!black}{$\Rightarrow$} CC|\makebox[6mm][r]{\nicefrac{C}{2}}    $\rightarrow$ \makebox[11mm][r]{\SI{1.91}{\volt}}
                                 \textcolor{black!50}{\textbf{>>}}         CV|\makebox[11mm][r]{\SI{1.91}{\volt}}  $\rightarrow$ \makebox[11mm][r]{\SI{8.0}{\amperehour}}              \\
                             &	 \textcolor{red}{$\Leftarrow$}             CC|\makebox[6mm][r]{1C}                 $\rightarrow$ \makebox[11mm][r]{\SI{1.000}{\volt}}                  \\
                \hdashline[0.5pt/1.5pt]
                \textrm{III} &	 \textcolor{green!50!black}{$\Rightarrow$} CC|\makebox[6mm][r]{\nicefrac{C}{2}}    $\rightarrow$ \makebox[11mm][r]{\SI{1.91}{\volt}}
                                 \textcolor{black!50}{\textbf{>>}}         CV|\makebox[11mm][r]{\SI{1.91}{\volt}}  $\rightarrow$ \makebox[11mm][r]{\SI{8.4}{\amperehour}}              \\
                             &   \textcolor{red}{$\Leftarrow$}             CC|\makebox[6mm][r]{1C}                 $\rightarrow$ \makebox[11mm][r]{\SI{1.000}{\volt}}                  \\
                \hdashline[0.5pt/1.5pt]
                \textrm{IV}  &	 \textcolor{green!50!black}{$\Rightarrow$} CC|\makebox[6mm][r]{\nicefrac{2C}{3}}   $\rightarrow$ \makebox[11mm][r]{\SI{5.0}{\amperehour}}
                                 \textcolor{black!50}{\textbf{>>}}         CC|\makebox[11mm][r]{\nicefrac{7C}{12}} $\rightarrow$ \makebox[11mm][r]{\SI{6.2}{\amperehour}}
                                 \textcolor{black!50}{\textbf{>>}}         CC|\makebox[6mm][r]{\nicefrac{C}{2}}    $\rightarrow$ \makebox[11mm][r]{\SI{1.91}{\volt}}
                                 \textcolor{black!50}{\textbf{>>}}         CV|\makebox[11mm][r]{\SI{1.91}{\volt}}  $\rightarrow$ \makebox[11mm][r]{\SI{8.4}{\amperehour}}              \\
                             &   \textcolor{red}{$\Leftarrow$}             CC|\makebox[6mm][r]{1C}                 $\rightarrow$ \makebox[11mm][r]{\SI{1.000}{\volt}}                  \\
                \hdashline[0.5pt/1.5pt]
                \textrm{V}   &	 \textcolor{green!50!black}{$\Rightarrow$} CC|\makebox[6mm][r]{\nicefrac{C}{2}}    $\rightarrow$ \makebox[11mm][r]{\SI{1.91}{\volt}}
                                 \textcolor{black!50}{\textbf{>>}}         CV|\makebox[11mm][r]{\SI{1.91}{\volt}}  $\rightarrow$ \makebox[11mm][r]{\SI{9.2}{\amperehour}}              \\
                             &   \textcolor{red}{$\Leftarrow$}             CC|\makebox[6mm][r]{1C}                 $\rightarrow$ \makebox[11mm][r]{\SI{1.525}{\volt}}
                                 \textcolor{black!50}{\textbf{>>}}         CC|\makebox[11mm][r]{\nicefrac{C}{2}}   $\rightarrow$ \makebox[11mm][r]{\SI{1.500}{\volt}}
                                 \textcolor{black!50}{\textbf{>>}}         CC|\makebox[6mm][r]{\nicefrac{C}{10}}   $\rightarrow$ \makebox[11mm][r]{\SI{1.000}{\volt}}                  \\
                \hdashline[0.5pt/1.5pt]
                \textrm{VI}  &   \textcolor{green!50!black}{$\Rightarrow$} CC|\makebox[6mm][r]{\nicefrac{2C}{3}}   $\rightarrow$ \makebox[11mm][r]{\SI{6.0}{\amperehour}}
                                 \textcolor{black!50}{\textbf{>>}}         CC|\makebox[11mm][r]{\nicefrac{7C}{12}} $\rightarrow$ \makebox[11mm][r]{\SI{7.2}{\amperehour}}
                                 \textcolor{black!50}{\textbf{>>}}         CC|\makebox[6mm][r]{\nicefrac{C}{2}}    $\rightarrow$ \makebox[11mm][r]{\SI{1.91}{\volt}}
                                 \textcolor{black!50}{\textbf{>>}}         CV|\makebox[11mm][r]{\SI{1.91}{\volt}}  $\rightarrow$ \SI{9.2}{\amperehour}                                 \\
                             &   \textcolor{red}{$\Leftarrow$}             CC|\makebox[6mm][r]{1C}                 $\rightarrow$ \makebox[11mm][r]{\SI{1.525}{\volt}}
                                 \textcolor{black!50}{\textbf{>>}}         CC|\makebox[11mm][r]{\nicefrac{C}{2}}   $\rightarrow$ \makebox[11mm][r]{\SI{1.500}{\volt}}
                                 \textcolor{black!50}{\textbf{>>}}         CC|\makebox[6mm][r]{\nicefrac{C}{10}}   $\rightarrow$ \makebox[11mm][r]{\SI{1.000}{\volt}}                  \\
                \bottomrule	
    	    \end{tabular}
    	\end{center}
    \end{table*}
    \par
    In the first two scenarios,
    the base assumptions are tested.
    Scenario \textrm{I} (cyan) uses the same cycling protocol as the reference simulation (blue),
    but uses a \SI{50}{\percent} reduced OER rate constant.
    Scenario \textrm{II} (green) does not change any reaction rate constants but examines whether the postulated advantage of intermediate C-rates has an observable influence.
    For this purpose,
    the charging current is reduced to \nicefrac{C}{2} as suggested by \cref{fig:reaction_competition}a,
    which then also needs an adaption of the upper voltage limit to reach the CV phase before the \SI{8}{\amperehour} limit is attained.
    It is reduced to \SI{1.91}{\volt}.
    The discharge procedure is untouched.
    \par
    In \cref{fig:optimisation_oer}a,
    we see that the cell voltage curves are not too different from the reference case for these two scenarios.
    Scenario \textrm{I} using the same cycling protocol is nearly congruent to the base scenario,
    and scenario \textrm{II} features a lowered charging branch due to the reduced charging speed.
    Visible differences arise in \cref{fig:optimisation_oer}b,
    where both scenarios have a higher Coulombic efficiency (\SI{99.80}{\percent} and \SI{99.72}{\percent}) than the reference's \SI{99.60}{\percent} (solid lines),
    which also reflects in a slower permanent conversion of \ce{ZnO} (dashed lines) to \ce{Zn} (dotted lines).
    These two quantities are depicted as volume-averaged volume fraction in the anode,
    evaluated at the end of charge,
    when \ce{ZnO} (and subsequent zincate) shortage has the most severe influence.
    %
    \par
    Based on this enhancement of the CE,
    we explore now whether this may be also translated to a gain in capacity by pushing the end-of-charge limit to \SI{8.4}{\amperehour}.
    Scenario \textrm{III} (olive) differs from scenario \textrm{II} just by changing this limit.
    Scenario \textrm{IV} (orange) modifies additionally the charging behaviour by trying to implement a step-wise change of charging rate to stay in an optimum as suggested by \cref{fig:reaction_competition}b.
    The discharge procedure is still untouched.
    \par
    The cell voltage curves in \cref{fig:optimisation_oer}a feature the enlarged transferred charge and in case of scenario \textrm{IV} the now staircase-shaped charging branch due to the step-wise switching of charging currents when specific amounts of charge have been transferred.
    The slightly augmented capacity lets the CE drop noticeably to around \SI{99.41}{\percent} for both as depicted in \cref{fig:optimisation_oer}b,
    below the projected minimum of \SI{99.5}{\percent}.
    Thus,
    the continued reduction of \ce{ZnO} volume fraction is accelerated compared to the base scenario,
    which also reflects in an accelerated degradation starting between cycles $130$ and $150$ as visible from the CE.
    \par
    %
    In a last step,
    the findings of \cref{fig:reaction_competition}c are integrated into the discharge phase of the cycling protocol.
    For this purpose,
    Scenarios \textrm{V} (purple) and \textrm{VI} (pink) are derived from \textrm{III} and \textrm{IV}, respectively,
    by adapting the discharge current to \nicefrac{C}{2} when \SI{1.525}{\volt} is underrun and subsequently to \nicefrac{C}{5} when cell voltage falls below \SI{1.5}{\volt}.
    For scenario \textrm{VI},
    the switching points for the charging current are slightly adapted to the new use-case,
    see \cref{tbl:protocols}.
    In the same step,
    the end-of-charge limit is again raised,
    now to \SI{9.2}{\amperehour},
    to make use of the anticipated capacity accessed by deep discharge.
    \par
    The alteration of the cycling protocols are clearly visible in the cell voltage curves of \cref{fig:optimisation_oer}a:
    Both scenarios have an important increase of the charging phase allowing more charge to be transferred,
    and the discharge branch shows the two switching points of discharge current with the resulting changes in cell voltage.
    Furthermore,
    the cycling to lower SOCs shifts the cell voltage curves down to lower voltages visible in the charge and discharge branches.
    For scenario \textrm{VI},
    the charging branch is again staircase-shaped due to the changes in charging currents.
    This changed cycling regime shows nearly as good efficiencies for both scenarios as for the reference simulation (\SI{99.52}{\percent} and \SI{99.56}{\percent}),
    see \cref{fig:optimisation_oer}b,
    with scenario \textrm{VI} being slightly closer.
    As usual this translates equally to the conversion of \ce{ZnO} to \ce{Zn} with progressing cycling,
    which happens with the same tendencies close to that of the base scenario.
    %
    \\~\\
    The scenarios show that a reduction of the OER rate as well as a considered choice of cycling protocol may improve Coulombic efficiency,
    despite the reference simulation's cycling protocol being already quite good.
    This may be translated into either an increased cycle life or greater capacity utilisation.
    \par
    Scenario \textrm{I} is easily explained: By keeping the standard cycling protocol the reduced reaction rate lowers the amount of charge being used for oxygen formation,
    rendering the whole process more efficient.
    This reduces the imbalance between charging and discharging operation,
    slowing down the conversion of \ce{ZnO} to \ce{Zn} and hence will delay the initiation of cell death by \ce{ZnO} and zincate shortage.
    As already mentioned,
    this type of OER suppression is linked to the use of expensive additives like cobalt and thus not always an option.
    Therefore,
    it is noteworthy that similar improvements in CE can be achieved through a thoughtful choice of the cycle protocol,
    as demonstrated by scenario \textrm{II}.
    The use of intermediate charging rates,
    as implied by \cref{fig:reaction_competition}a,
    indeed has a beneficial effect on the competition between OER and proton insertion/extraction reaction.
    As indicated by the development of the volume-averaged \ce{Zn} and \ce{ZnO} curves in \cref{fig:optimisation_oer}b,
    this will also lead to an extended cycle life.
    %
    \par
    The examination of scenario \textrm{III} reveals that these gains in efficiency and thus cycle life are only possible in a narrow corridor of arrangements.
    Enhancing capacity in particular means walking a fine line between improvement and deterioration.
    Increasing the utilised capacity by \SI{5}{\percent} to \SI{8.4}{\amperehour} extents the accelerated oxygen formation period to such lengths that the CE drops by approximately \SI{0.3}{\percent},
    which is already enough to approximately halve the cycle life and thus not justifying the modest increase in capacity.
    The attempt to overcome this strong reduction of CE with scenario \textrm{IV} by using optimal charging rates as suggested by \cref{fig:reaction_competition}b does not yield any change in the performance metrics.
    \par
    But the integration of adapted discharge C-rates as discussed on ground of \cref{fig:reaction_competition}c proves that larger capacities may be accessed.
    Both scenarios,
    \textrm{V} and \textrm{VI},
    show nearly as good CE as the reference simulation,
    with \textrm{VI} and its adapting charging rates being slightly ahead.
    This might be surprising having the nearly identical outcome of scenarios \textrm{III} and \textrm{IV} in mind,
    but for this scenario the switching points for charging currents were adapted to the deep discharge in this scenario.
    These new switching points seem to match better the predicted OER suppressing behaviour in \cref{fig:reaction_competition}b.
    At the same time,
    both scenarios enable the \ce{Ni}/\ce{Zn} cell to use \SI{15}{\percent} more capacity by only conceding a small amount of cycle life.
    The larger part of this improvement is possible due to decreasing the lower SOC limit from the reference case's approximately \SI{9.5}{\percent} (cf. \cref{fig:longterm}c) to now approximately \SI{2.3}{\percent}.
    \par
    Generally,
    we see that a careful combination of cycling operations allows to access additional performance of the cell,
    if lower and changing C-rates are acceptable in the cell's use-case.
    The exemplary scenarios showed that this may be used for extended cycle life or larger capacity usage.
    But also combinations of both are possible,
    e.g. by using scenario \textrm{VI} and increasing the capacity by only \SI{12.5}{\percent} which would translate to a cycled capacity of \SI{9.0}{\amperehour} and an efficiency of about \SI{99.65}{\percent}.
    Furthermore,
    a even more thoroughly chosen set of switching points for the charge and discharge C-rates of scenarios \textrm{IV} and \textrm{VI} could exploit the potential of the optimisation approaches presented here even better.
    The more easily accessible capacity seems to be that of the lower SOC range,
    where nearly \SI{1}{\amperehour} are added in case of scenario \textrm{VI} compared to only \SI{0.2}{\amperehour} in the upper SOC range.
    \par
    These rather theoretical findings should in principle be applicable to physical \ce{Ni}/\ce{Zn} cells as well.
    While finding e.g. optimal switching points for C-rates could be a tedious work and exploiting deep discharge regions too much could lead to e.g. cracking of the \ce{Ni} electrode due to volume expansion,
    leading to a reduced cycle life,
    the careful exploration of these cycling protocol optimisation strategies may increase the capacity and/or cycle life by several percent,
    making \ce{Ni}/\ce{Zn} cells more competitive.
    %
    %
    %
  %
  %
  %
%
%
%
\section*{Conclusion}
\label{sec:conclusion}
  A physico-chemical and volume-averaged three-dimensional model of a \ce{Ni}/\ce{Zn} cell with additional focus on side reactions and active material particles was presented.
  The former includes,
  besides the important OER and HER,
  a rudimentary implementation of \ce{O2} and \ce{H2} outgassing and \ce{O2} recombination with \ce{Zn}.
  The latter comprises hindered transport due to \ce{ZnO} precipitation around \ce{Zn} particles as well as solid diffusion of inserted \ce{H^+} in \ce{Ni} active material.
  \par
  The model has been adjusted to experimental data of a prototypical \ce{Ni}/\ce{Zn} cell,
  and computational studies have been conducted to compare the behaviour of the virtual twin.
  Through 3D simulation and comparison to $\mu$-XRF data of the prototypical cell, it has been shown that the model is capable of reproducing the zinc shape change effect.
  Analysing the convective flow throughout a cycle revealed that density changes of the electrolyte solution due to variations in the hydroxide and zincate concentration surpass pore volume changes due to \ce{Zn} dissolution/deposition and \ce{ZnO} precipitation/dissolution.
  Hence,
  electrolyte solution is counter-intuitively pressed out of the \ce{Zn} electrode during charging while it is sucked back in during discharge.
  This pattern may assist the zinc shape change through convecting zincate in waves towards the centre of the electrode,
  as it already was pointed out by experiments in literature.
  \par
  Analysis of long-term cycling simulations revealed that \ce{Zn}/\ce{ZnO} imbalance in each cycle is a primary cause of cell failure at low- and mid-range cycle numbers.
  This imbalance,
  which manifests as insufficient Coulombic efficiency,
  stems from progressive depletion of \ce{ZnO} and ultimately \ce{Zn(OH)4^{2-}}.
  The underlying cause is the oxygen evolution reaction (OER),
  which becomes increasingly active at higher states of charge.
  \par
  Based on these insights,
  two key optimisation strategies for extending cycle life have been identified.
  The first approach involves using additives such as cobalt to manipulate the standard reduction potential of the OER,
  thereby slowing the detrimental reaction.
  Alternatively,
  the cycling protocol may be modified to exploit an observed C-rate dependence of the competition between insertion and oxygen evolution reactions.
  This second strategy offers the attractive possibility of avoiding additives like \ce{Co} by optimising the cycling protocol,
  requiring only careful selection of appropriate C-rates and usable capacity ranges.
  As demonstrated,
  this approach is also suitable to boost the cycle life and cycled capacity considerably.
  %
  %
  %
%
%
%

\clearpage

\section*{Author Contributions}

\textbf{F. K. Schwab:} Conceptualisation, Methodology, Software, Visualisation, Validation, Formal analysis, Writing - Original Draft, Writing - Review \& Editing
\textbf{B. Doppl:} Conceptualisation, Methodology, Writing - Review \& Editing
\textbf{N. J. Herrmann:} Conceptualisation
\textbf{A. Boudet} Investigation, Validation, Writing - Review \& Editing
\textbf{S. Mirhashemi} Investigation, Validation, Funding acquisition, Writing - Review \& Editing
\textbf{S. Brimaud} Investigation, Validation, Funding acquisition, Writing - Review \& Editing
\textbf{B. Horstmann:} Conceptualisation, Supervision, Funding acquisition, Writing - Review \& Editing

\section*{Electronic Supplementary Information (ESI)\dag}
The authors have cited additional references within the Electronic Supplementary Information\dag.\cite{
EoECPS2023,
Newman2021,
Schmitt2020,
Bolay2022,
SunErgyInternal2023,
SunErgyInternal2024,
CergyUniversity2024,
ZSWInternal,
Liu1981a,
Horn2003,
Matula1979,
Miller1941,
Meaden1965,
Orlov1996,
Freund1980,
Nishi2012,
Pierson1996,
Motupally1995,
Bhatia1968,
May1978,
Dean1999,
Liu1981b,
Barthel2003,
Atkins2006,
See1997,
Katchalsky1965,
Fong2020,
Malaie2023,
Sunu1978,
Bennion1964,
Usanovich1951,
Gilliam2007,
Dyson1968,
Lang1956,
Himy1986,
Trudgeon2019,
Takahashi1974,
Kriegsmann1999,
See1998,
Falk1969,
McBreen1967,
Dirkse1970,
Gendler1976,
Nanis1970,
DeWane1968,
Trudgeon2020,
Akerlof1941,
Yushkevich1967,
Lengyel1963,
Cussler2009,
Tham1970,
Hodges2023,
Kube2021,
Kolchko1959,
Zaytsev1992,
Guo2010,
Siu1997,
Jackson1986,
Bear1972,
Kozeny1927,
Carman1937,
Feldkamp1969,
Blazy2021,
Yeo1985,
Leverett1941,
Danner2014,
Foroughi2022,
Schmitt2020thesis,
Watanabe1997,
Watanabe1998,
Bockris1972,
Bard2000,
Lide2005,
Hendrikx1984,
Jain1992,
Moore1982}

\section*{Acknowledgements}
This work is supported by the European Union’s Horizon 2020 research and innovation programme under grant
agreement no 963576 (LoLaBat) and by the Franco-German project ZABSES,
co-funded by the Agence Nationale de la Recherche (ANR, ANR-22-MER3-0005-02) and the Federal Ministry of Education and Research of Germany (BMBF, 03XP0505A).
The authors acknowledge support by the state of Baden-Württemberg through bwHPC
and the German Research Foundation (DFG) through grant no INST 40/575-1 FUGG (JUSTUS 2 cluster).
Emanuele Marini (ZSW) is acknowledged for the acquisition of the $\mu$-XRF experimental data.

\section*{Conflict of Interest}
There are no conflicts to declare.



\setlength{\bibsep}{0.0cm}
\bibliographystyle{Wiley-chemistry}
\bibliography{dlr.bib,ESI_bibliography.bib}

\end{document}